\documentclass[11pt,a4wide]{article}
\usepackage{graphicx}
\usepackage{lscape}
\usepackage{color}
\usepackage[british]{babel}
\usepackage{amssymb,amsmath,amsfonts,mathrsfs,verbatim,xkeyval,bm}
\usepackage{textcomp}
\usepackage{fancyhdr}

\definecolor{mygray}{RGB}{194,194,194} 
\definecolor{myblue}{RGB}{0,138,230} 
\definecolor{mypink}{RGB}{224,11,122} 
\definecolor{mydarkgray}{RGB}{61,61,61} 
\definecolor{mylightgray}{RGB}{128,128,128} 
\definecolor{myverylightgray}{RGB}{194,194,194} 
\definecolor{myyellowgreen}{RGB}{122,224,11}
\definecolor{mypurple}{RGB}{122,11,224}

\def\build#1_#2^#3{\mathrel{
    \mathop{\kern 0pt#1}\limits_{#2}^{#3}}}

\definecolor{myblue}{RGB}{0,138,230} 
\definecolor{mypink}{RGB}{224,11,122} 
\definecolor{mygray}{RGB}{194,194,194} 
\definecolor{mydarkgray}{RGB}{61,61,61} 
\definecolor{myyellowgreen}{RGB}{122,224,11}
\definecolor{mypurple}{RGB}{122,11,224}
\newtheorem{theorem}{Theorem}[section]

\newtheorem{bookrule}[theorem]{Rule}

\begin{document}
\begin{center}
{\Large\bf Secondary resonances and the boundary of effective stability 
of Trojan motions}\\
\vskip 1cm
Roc\'{i}o Isabel P\'{a}ez$^1$ and Christos\,Efthymiopoulos$^1$\\
$^1$Research Center for Astronomy and Applied Mathematics, 
Academy of Athens\\
\vskip 1cm
\end{center}

\noindent
{\small {\bf Abstract:} 

One of the most interesting features in the libration domain of 
co-orbital motions is the existence of secondary resonances.  For some
combinations of physical parameters, these resonances occupy a large
fraction of the domain of stability and rule the dynamics within the
stable tadpole region. In this work, we present an application of a 
recently introduced `basic Hamiltonian model' $H_b$ for Trojan dynamics 
\cite{PaezEfthy2015},~\cite{Paezetal2016}: we show that the inner border 
of the secondary resonance of lowermost order, as defined by $H_b$, 
provides a good estimation of the region in phase-space for which the 
orbits remain regular regardless the orbital parameters of the system. 
The computation of this boundary is straightforward by combining a 
resonant normal form calculation in conjunction with an `asymmetric 
expansion' of the Hamiltonian around the libration points, which speeds 
up convergence. Applications to the determination of the effective 
stability domain for exoplanetary Trojans (planet-sized objects 
or asteroids) which may accompany giant exoplanets are discussed.

\section{Introduction}\label{sec:intro}

Despite the theoretical possibility of the existence of Trojan
exoplanets (\cite{LaughlinChamb-02},
\cite{Beugetal-07},~\cite{CressNel-09}), no such body has been
identified so far in exoplanet surveys.  This lack of identification
may reflect formation constrains, constrains to detectability
(\cite{Haghietal-13},~\cite{Dobro-13}, \cite{Leleu-15},
\cite{Leleu-thesis}, \cite{Leleu-17}), or it may simply be due to
stability reasons.  In this framework, the question of `effective
stability', i.e.  stability of the orbit of a Trojan body for times as
long as a considerable fraction of the age of the hosting system,
comes to the surface. The question of effective stability has been
addressed nearly exhaustively in the case of Trojan asteroids in our
Solar System (see, for example,~\cite{Milani-93}, \cite{Levetal-97},
\cite{GiorSko-97}, \cite{Tsig-05}, \cite{RobGab-06},
\cite{LhotEfthyDvo-08}, \cite{Dvoetal-10}, \cite{Lyketal-11}) from
both numerical and analytical approaches, but only scarcely in the
case of exoplanetary systems (see~\cite{Nauen-02}, \cite{ErdiSan-05},
\cite{Schwarzetal-09}, \cite{Efthy-13}). One main reason for the
scarcity of results in this latter case is the vast volume of
parameter space to be investigated, in conjunction with the multi-body
nature of the problem: to determine the long-term stability of Trojan
motions becomes essentially a problem of secular dynamics with as many
degrees of freedom as the number of planets in the system under
consideration.  Any attempt to face the problem other than numerical
simulation clearly requires a simplification of the dynamical model,
without this leading to oversimplified conclusions regarding the
long-term orbital stability.

In the present work, we discuss a key property of the dynamics
  induced by \emph{secondary resonances} in the domain of Trojan
  motions, which in addition to its own proper interest, can serve
  also the purpose of obtaining a simple analytical
  estimate of the effective stability boundary of Trojan motions in
  hypothetical exoplanetary systems.  Our analysis of the resonant
dynamics stems from a set of considerations or assumptions, whose
validity can be most easily judged by comparison with some results and
figures of a previous work of ours (\cite{PaezEfthy2015}) as
follows:\\ 1) In~\cite{PaezEfthy2015} we provided a formalism of the
problem of the dynamics of Trojan bodies in the Hamiltonian context,
which recovers all essential features as discovered in previous
literature (\cite{Erdi-88}, \cite{Erdi-97}, \cite{Morais-99},
\cite{Morais-01}); preceding works, however, focus mostly on a direct
investigation of the equations of motion, averaged or not with respect
to short period terms. In our works we stressed, instead, a main
advantage of the new formalism, namely the allowance to recruit the
full machinery of Hamiltonian methods in order to better analyze the
problem under study.

2) We investigated the features of Trojan dynamics which hold under
three physically relevant assumptions: i) that the motions of all
planets, including the Trojan body, are close to planar, ii) that the
Trojan body is small enough to be considered as test particle, in
agreement with formation scenaria which suggest that exo-Trojans should
be at most Mars-sized objects (\cite{Beugetal-07}), and iii) that the secular
dynamics of the hosting system is such that the eccentricity vector of
the primary companion of the Trojan body undergoes circulation with a
nearly constant frequency $g'$, and has a length which undergoes
variations around some non-zero value $e_0$.

Under assumptions (i) to (iii), we find that the Hamiltonian of motion 
of the Trojan body, averaged over short period terms for the motions of 
the remaining planets, can be decomposed in the form $H = H_b + H_{sec}$ 
where: $H_b$, called the `basic model', describes short period and synodic 
motions, and yields a constant proper eccentricity for the Trojan body, 
and $H_{sec}$ contains all remaining secular perturbations.
Furthermore $H_b$ has a universal form, i.e., it suffices to redefine
the physical meaning of the angular canonical variables, to keep 
its form unaltered in the whole hierarchy of restricted problems 
(circular, elliptic, secular with more than one perturbing planets).

3) The decomposition $H=H_b+H_{sec}$ leads to a specific physical
  understanding of the dynamics when the primary planet has a mass in
  the giant planet range. In this case, the three timescales related
  to the short-period, synodic and secular motions have a separation
  by about one or less order of magnitude from each other. Then, due
to the specific features of $H_b$ described above, we arrive at the
following key remark: the model $H_b$ produces, in phase space, a set
of secondary resonances corresponding to commensurabilities between
the frequencies of the short-period and synodic
motions~(\cite{Erdietal-07}). It is easy to see that these are the
only secondary resonances which occupy a non-zero volume in phase
space in the whole hierarchy of restricted problems that one could use
as dynamical models for the Trojan body. However, there exists a {\it
  modulation} effect (\cite{Chirikov-85}) due to the influence of
$H_{sec}$ on this set of secondary resonances: the separatrices
pulsate slowly (with one or more secular frequencies) and, as a
result, in the `domain of uncertainty' (\cite{Neishtadt-87}) created
by such pulsations, the motions become chaotic. Such an effect is
possible to visualize already in the Elliptic Restricted Three-Body
Problem, namely the simplest model with non trivial $H_{sec}$. The
reader is refered to Figures 5 to 15 of~\cite{PaezEfthy2015} which
show in detail the statements below, by exemplifying the outcome of
the modulation effects for the secondary resonances 1:5 up to 1:12,
when the modulus of the eccentricity vector $e_0$ of the primary
companion varies from $e_0=0$ to just a moderate value $e_0=0.1$. By
inspecting the stability maps in the space of the Trojan body's proper
elements, one sees that, for $e_0$ slightly larger than zero, the
separatrix pulsation for the secondary resonances becomes large enough
so as to wipe out nearly completely the domain of stable motions
occupied by such resonances. As a result, the only remaining stable
motions are those in a inner (closer to the libration center) domain
devoid of secondary resonances. In fact, as found in many works
(e.g.~\cite{RobGab-06}, \cite{MarzSch-07})
there can still be resonances involving one
or more secular frequencies which penetrate this innermost stability
domain. However, since these resonances are thin and typically do not
overlap, they can only induce a very slow chaotic diffusion of the
Arnold type, which can be neglected for all practical purposes.
Hence, the innermost domain,
devoid of the secondary resonances of $H_b$, meets all criteria of
effective stability, and, indeed, stability maps indicate the
robustness of this domain against variations of the orbital parameters
of the Trojan body.

\subsection{Summary of the method}
Stemming from remarks (1) to (3) above, we propose below a practical 
method to define the effective stability domain of Trojan motions. 
This is based on the following steps: \\
{\it Step 1:} analyze a given system where hypothetical Trojan bodies 
are sought for and compute 
the Hamiltonian $H_b$, \\
{\it Step 2:} identify the largest in size (typically lowest in order) 
secondary resonance of $H_b$ for given parameter values,\\
{\it Step 3:} compute a resonant normal form and evaluate the 
theoretical separatrices of the identified secondary resonance,\\
{\it Step 4:} assume that all stable domains of resonant motions 
extending beyond the innermost (closest to the libration center) 
branch of the theoretical separatrices were wiped out by secular 
modulation effects.

Then, the locus $S$ formed by the intersection of the family of all 
computed innermost theoretical separatrices, along the dominant 
secondary resonance, with any chosen (with respect to phases) plane 
$PP$ of Trojan proper elements, yields the boundary of the effectively 
stability domain in the plane $PP$.    

The above computation is fast and straightforward 
to perform with modern computer algebra programs, and thus competitive 
to large grid computations of effective stability maps.
In our own implementation we use the normal form method adopted in
\cite{PaezLocat2015},~\cite{Paezetal2016}.

 In the rest of the paper, we discuss both the dynamical role of
  the secondary resonances in delimiting the main domain of effective
  stability as well as our particular analytical method of computing
  the border of this domain.  The structure of the paper is as
  follows: In Section 2, we review the derivation and features of
the Hamiltonian $H_b$, as well as our way to expand $H_b$ in a form
form suitable for resonant normal form computations. A novel feature
is the adoption of an `asymmetric expansion' which improves
convergence. Section 3 explains in detail the realization of the {\it
  Steps 1-4}, in particular the computation of the theoretical
separatrices of the dominant secondary resonance and their
superposition to stability maps in the space of proper elements. 
  Section 4 contains the main results: (a) we provide numerical
  evidence, based on stability maps, of how the separatrices of the
  secondary resonances of the 'basic model' $H_b$ act as delimiters of
  the effective stability domain; (b) we use an analytical method to
  estimate this boundary; (c) we discuss the robustness of the present
  approach against changing the model's parameters (masses and
  eccentricities), as well as when considering, in the numerical
  integrations, the full three-body problem instead of the ERTBP.
Section 5 summarizes our main conclusions.

\section{Basic Hamiltonian $H_b$ and its asymmetric expansion}\label{sec:asymexp}

\subsection{Main features of the basic model $H_b$}\label{sbs:hb}

In~\cite{PaezEfthy2015}, a Hamiltonian formulation was provided for the 
Trojan motion which applies to the planar Elliptic Restricted Three-Body 
Problem (ERTBP) with a central mass, a primary perturber or simply `primary', 
and the Trojan test particle, or when $S$ additional perturbing bodies are 
present but far from MMRs, the so-called 'Restricted Multi-Planet Problem' 
(RMPP)). The Hamiltonian reads
\begin{equation}\label{eq:h_rmpp}
H = H_b\,(Y_f,\phi_f,u,v,Y_p;\mu,e'_0)  
+\, H_{sec}\,(Y_f,\phi_f,u,v,Y_p,\phi,Y_1,\phi_1,\ldots,Y_S,\phi_S)~~.
\end{equation}
In Eq.~\eqref{eq:h_rmpp}, the variables $(\phi_f,Y_f)$, $(u,v)$ and
$(\phi,Y_p)$ are pairs of action-angle variables, whose definition
stems from Delaunay-like variables following a sequence of four
consecutive canonical transformations (see Appendix A). In
particular, $(Y_f,\phi_f)$ are action-angle variables describing the
fast degree of freedom, of frequency
\begin{equation}\label{eq:fast_freq}
\omega_f \equiv \dot{\phi}_f = 1 - \frac{27}{8}\, \mu + g' + \ldots~~,
\end{equation}
where $g'$ is fundamental frequency of precession of the primary's 
perihelion. The pair $(u,v)$ describe the particle's synodic librations, 
$u\simeq \lambda-\lambda'-\pi/3$, $v\simeq\sqrt{a}-1$, with $\lambda$, 
$\lambda'$ the mean longitudes of the test particle and of the primary, 
$a$ the particle's major semi-axis, and $a'=1$. The associated frequency 
at the libration center is  
\begin{equation}\label{eq:sin_freq}
\omega_s \equiv \dot{\phi}_s = - \sqrt{\frac{27 \mu}{4}} + \ldots ~~.
\end{equation}
Finally, the secular motion of the test particle's eccentricity vector
$\left( e\cos(\omega-\omega'),e\sin(\omega-\omega') \right)$, where $e$
is the eccentricity and $\omega$,$\omega'$ are the arguments of the
perihelion of the test particle and the primary respectively, is
described by a circulation around the forced equilibrium point, given
in our variables by a set of action angle variables $(Y_p,\phi)$. The
associated secular frequency is
\begin{equation}\label{eq:sec_freq}
g \equiv \dot{\phi} = \frac{27}{8}\, \mu - g' + \ldots~~.
\end{equation}
We call the term $H_b$ in the Hamiltonian of Eq.~\eqref{eq:h_rmpp} the 
`basic Hamiltonian model' for Trojan motions in the 1:1 MMR. Its detailed 
form is given in the Supplementary Online Material of~\cite{PaezEfthy2015}. 
We find
\begin{equation}\label{eq:hbasic}
H_b = -\frac{1}{2(1+v)^2}-v +(1+g')Y_f - g' Y_p - \mu {\cal F}^{(0)}
(u,\phi_f,v,Y_f-Y_p;e_0')~~.
\end{equation}
The physical parameters entering into $H_b$ are i) the mass parameter
$\mu=\frac{m'}{m'+M}$, where $M$ is the mass of the central mass and
$m'$ the mass of the primary, ii) the mean value of the length of the
eccentricity vector of the heliocentric orbit of the primary perturber,
$e'_0$. In the ERTBP, one has simply $e'=e_0'$, $g'=0$ (implying also
$\omega' \equiv \mathrm{const}$). However, the form of $H_b$ remains
the same in both the ERTBP and the RMPP. 
In particular, the angle $\phi$ is defined via a `shift transformation'
depending only on the relative difference $\Delta\omega=\omega-\omega'$. 
Physically, the secular dynamics
induced under $H_b$ appears the same in the ERTBP and in the
RMPP, when, in the latter case, it is 
viewed in apsidal co-rotation with the
primary. 
Furthermore, since the angle $\phi$ is ignorable in $H_b$,
the action variable $Y_p$ is an integral of the basic
Hamiltonian. Then, the ERTBP and the RMPP are diversified only by
their different form of the functions $H_{sec}$.  In particular, in
the RMPP case $H_{sec}$ contains also pairs of action angle variables
associated with the secular precessions of the S additional bodies,
while in the ERTBP it contains only the angle $\phi$ associated with
the secular precession of the test particle.  Finally, $H_{sec}$
disappears all together in the circular RTBP. Thus, $H_b$ becomes the exact
Hamiltonian in this case. Note, however, that in the ERTBP $H_b$ is
{\it not} equal to the ensemble of all terms independent of $e'$. 

The basic model $H_b$ represents a drastic reduction of the number of degrees 
of freedom with respect to the original problem. In the sequel, 
we will focus on one particular feature of $H_b$, namely the presence 
of {\it secondary resonances}, which correspond to commensurability 
relations between $\omega_f$ and $\omega_s$. In particular, we focus
on the role of these resonances in practically determining the boundary of the 
effective domain of stability for the Trojan motions.

\subsection{Asymmetric expansion}\label{sbs:asymmexp}
The resonant normal form computed in Section 3 below provides a model 
for studying the dynamics within or near a secondary resonance of the
form:
\begin{equation}\label{eq:main_sec_res}
m_f\omega_f+m_s\omega_s = 0~.
\end{equation}
A {\it non-resonant} normal form for the model $H_b$, allows to find
the location of secondary resonances in a space of suitably defined
proper elements for the Trojan body (see~\cite{Paezetal2016}).
However, the non-resonant normal form does not allow to compute the
local phase portrait, i.e., the separatrices associated with each
resonance. Furthermore, all series expansions which are polynomial in
the variables $u,v$ exhibit poor convergence, a fact associated with
the singularity (collision with the primary) at $u=-\pi/3$. In order
to deal with this problem, a {\it partially expanded} version of the
$H_b$ can be used~\cite{PaezLocat2015}, in which all the powers of the
quantity $\beta(\tau)=\frac{1}{\sqrt{2-2\cos\tau}}$ (with $\tau =
u+\pi/3$) are kept unexpanded. This leads to a Hamiltonian of the
form {\small
\begin{equation} 
\begin{aligned}\label{eq:hb_with_beta_calY} 
H_b (&v,{\cal Y},\tau,\phi_f,Y_p) = -v
+ \sum_{i=0}^{\infty} \,(-1)^{i-1}(i+1)\,
\frac{v^i}{2}\, +\, {\cal Y} \,+ \,Y_p \\ +&
\,\mu \sum_{\substack{ m_1,m_2,m_3\\ k_1,k_2,k_3,j}}
a_{m_1,m_2,m_3,k_1,k_2,j}\, e'^{k_3} v^{m_1} \, \cos^{k_1}(\tau) \, 
\sin^{k_2}(\tau) {\cal Y}^{m_4} \,
\cos^{m_2} \phi_f \, \sin^{m_3} \phi_f \,
\beta^j(\tau)~~,
\end{aligned}
\end{equation}}
where $\tau = u + \pi/3$,
${\cal Y} = Y_f - Y_p$, and $a_{m_1,m_2,m_3,k_1,k_2,j}$ are rational 
numbers.
 
The librations in $\tau$ (or $u$) are represented in terms of the
synodic angle variable $\phi_s$, i.e., the phase of the synodic
libration.  The computation of a resonant normal form requires to
explicitly Fourier expand the terms of $H_b$ in \emph{both} angles
$\phi_f$ and $\phi_s$.  Although the
Hamiltonian~\eqref{eq:hb_with_beta_calY} represents a Fourier
expansion for the fast d.o.f. (angle $\phi_f$), there still remain the
powers of $\beta$ that must be expanded in powers of $u$ in order to
obtain a complete Fourier expansion in the angle $\phi_s$ as well. Due
to the singularity at $\tau=0$ (or $u=-\pi/3$), any Taylor expansion
of the functions $ \beta(\tau)^N = \frac{1}{\left( 2-2\cos\tau
  \right)^{N/2}}$, with $N\in \mathbb{N}$, around a certain $\tau_0$
is convergent only in the domain $\mathscr{D}_{\tau_0,\delta}$
centered at $\tau_0$ and of radius $\delta= \mathrm{Min}\{\tau_0,
2\pi-\tau_0\}$.  The most common approach consists of Taylor
expansions around the libration equilibrium point, located at
$\tau_0=\frac{\pi}{3}$, for L4, or $\tau_0=\frac{5\pi}{3}$ for L5.
The corresponding $\delta$ in this case is $\frac{\pi}{3}$. One finds
that many Trojan orbits, and important secondary resonances, may cross
this domain. In such cases, the resonant normal form construction is
obstructed by the poor convergence of the original Hamiltonian
expansion.

In order to face this problem, we find a different polynomial representation 
of the Hamiltonian $H_b$ in the variables $(u,v)$ by performing an 
\emph{asymmetric} expansion, i.e. expansion around a \emph{non-equilibrium 
point} $\tau_0\neq \pi/3$, selected to be further away from the singularity 
but close enough to the libration point, so that a re-ordering of the 
expansion in powers of $u$ yields a negligible term linear in $u$
(since $u = 0$ represents the equilibrium point of $H_b$). 
Here we choose $\tau_0 = \frac{\pi}{2}$. In this case, we obtain a 
polynomial expansion of the Hamiltonian in powers of the quantity 
$(\tau-\pi/2)$. Re-ordering the terms, we express it as a polynomial
in powers of $u$. It is immediate to see that any finite 
truncation of this expression yields a different polynomial than the 
one obtained by a finite truncation of the direct Taylor expansion 
around $\tau=\pi/3$. However, the new expression better represents 
the quantities $\beta(\tau)^N$ in a domain extended up to $\tau \sim \pi$. 
We call the expansion around $\tau_0 = \frac{\pi}{2}$ \emph{asymmetric}, 
while the one around $\tau_0 = \frac{\pi}{3}$ \emph{symmetric}.

Figure~\ref{fig:thebetas.png} shows the benefits of the asymmetric 
expansion when compared to the symmetric one. We consider the functions
\begin{equation}
\begin{aligned}\label{eq:the_betas}
B_{1}(\tau) & = \frac{\cos \tau}{\beta(\tau)} 
= \frac{\cos \tau}{(2-2\cos\tau)^{1/2}}~,
& B_{3}(\tau)  = \frac{\cos \tau}{\beta^3(\tau)} 
= \frac{\cos \tau}{(2-2\cos\tau)^{3/2}}~,\\
B_{5}(\tau) & =\frac{\cos \tau}{\beta^5(\tau)} 
= \frac{\cos \tau}{(2-2\cos\tau)^{5/2}}~,
& B_{7}(\tau)  = \frac{\cos \tau}{\beta^7(\tau)} 
= \frac{\cos \tau}{(2-2\cos\tau)^{7/2}}~,
\end{aligned}
\end{equation}
which represent the most common terms in powers of $\beta(\tau)$ appearing 
in Eq.~\eqref{eq:hb_with_beta_calY}. 
The symmetric Taylor expansion of $B_{1}$, $B_{3}$, $B_{5}$ and $B_{7}$ 
around $\tau_0= \pi/3$ yield the polynomials
\begin{equation}\label{eq:the_betas_sym}
B_{M,\pi/3}(u) = B_{M}(\pi/3) +  B^{(1)}_{M}(\pi/3) \, u + 
\frac{1}{2} B^{(2)}_{M}(\pi/3) \, u^2 + 
\frac{1}{6} B^{(3)}_{M}(\pi/3) \, u^3 + \ldots~,
\end{equation}
where $B^{(n)}_{M}(\pi/3)$ is the $n$-th derivative of the function $B_{M}$, 
evaluated at $u=\pi/3$, $M = 1,3,5,7$  and
$u = \tau - \pi/3$.
On the other hand, the asymmetric Taylor expansions of the same functions 
around $\tau_0 = \pi/2$ yield the polynomials
{\small
\begin{equation}\label{eq:the_betas_asym}
B_{M,\pi/2}(u)  = B_{M}(\pi/2) +  
B^{(1)}_{M}(\pi/2) \, (u-\frac{\pi}{6}) + 
\frac{1}{2} B^{(2)}_{M}(\pi/2)\, (u-\frac{\pi}{6})^2 + 
\frac{1}{6} B^{(3)}_{M}(\pi/2)\, (u-\frac{\pi}{6})^3 + \ldots~.
\end{equation}}

 \begin{figure}[h]
 \includegraphics[width=0.99\textwidth]{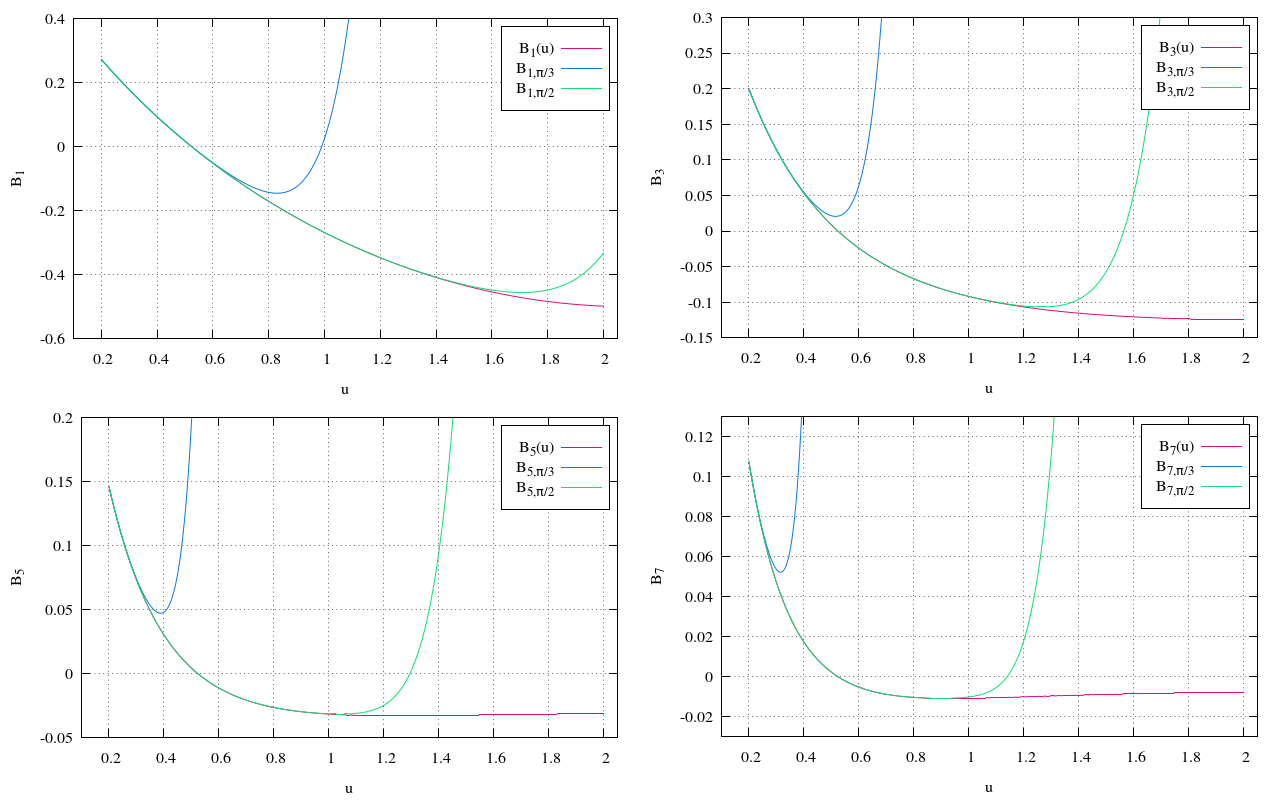}
 \caption{Comparison of the functions $B_{M}(u)$ (pink), with the
   expansions $B_{M,\pi/3}(u)$ (blue) and $B_{M,\pi/2}(u)$ (green), up to
   order 10 in $u$, for $M = 1,\,3,\,5,\,7$ and $u \in [0.2,2]$.}
 \label{fig:thebetas.png}
 \end{figure}

Fig.~\ref{fig:thebetas.png} compares the graphs of the original functions 
$B_{M}(u)$ (pink) with the two corresponding expansions $B_{M,\pi/3}(u)$
(symmetric, blue) and $B_{M,\pi/2}(u)$ (asymmetric, green) up to order 
10 in $u$. We see that both expansions provide a good representation of 
the original function up to a certain extent in $u$, but for increasing 
values of $u$, the asymmetric expansions are more accurate than the 
symmetric ones in a domain extending to higher values of $u$. The improvement 
in accuracy is more notorious as $M$ increases. In fact, we find that the 
polynomial approximations to $H_b$ found by the asymmetric expansion is 
accurate up to $u \sim 1 \,\mathrm{rad}$, which is enough to cover the 
effective stability domain in most physically relevant parameter values.
Only very close to $u\sim 0$, the symmetric expansion is marginally more 
accurate than the asymmetric one. This yields a slight shift of the 
equilibrium point of $H_b$ with respect to $u=0$, typically of about 
$\sim 10^{-8}$ rad, i.e. practically negligible. 

Similar results are found for the asymmetric and symmetric expansions 
of the functions $\frac{\sin \tau}{\beta(\tau)}$ appearing in $H_b$.
Finally, the asymmetric expansions for both types of functions can 
be easily performed by a closed set of formulas, given in the Appendix B.  

\section{Resonant normal form}

\subsection{Hamiltonian preparation}\label{sbs:impl}

The construction of the asymmetrically expanded $H_b$ consists of two 
steps: i) replacement of the expansions for the variables $(u,v)$ 
in Eq.~\eqref{eq:hb_with_beta_calY}, and ii) transformation to 
action-angle variables. 

In order to replace the polynomial truncations for the functions of
Eq.~\eqref{eq:the_betas} into Eq.~\eqref{eq:hb_with_beta_calY}, we
adopt the asymmetric expansion~\eqref{eq:the_betas_asym}, using
the formul\ae~provided in the Appendix B. Regarding $v$, it is
enough to consider the Taylor expansion of $H_b$ with respect to $v$
around zero. The maximum truncation order is determined in terms of a
`book-keeping parameter' (\cite{EfthyLaPlata}; see below).  After
these replacements, the Hamiltonian $H_b$ takes the form
\begin{equation}\label{eq:hb_asym_expand}
H_b (v,{\cal Y},u,\phi_f,Y_p) = 
Y_p + \sum_{\substack{m_1,m_2,\\m_3,m_4}}
\mathtt{a}_{m_1,m_2,m_3,m_4}\,  v^{m_1} \, u^{m_2}\, 
(\sqrt{{\cal Y}})^{m_3}\, \cos (m_4 \phi_f)~,
\end{equation}
where the real coefficients $\mathtt{a}_{m_1,m_2,m_3,m_4}$ depend on 
the parameters $\mu$ and $e_0'$ (or simply $e'$ in the ERTBP).

Next, we diagonalize $H_b$ in order to obtain a harmonic oscillator 
quadratic part for the synodic degree of freedom. Diagonalization is 
performed by the linear canonical transformation $(u,v)\rightarrow(U,V)$ 
defined by the set of formulas:
\begin{equation}\label{eq:fromuvtoUV}
\begin{pmatrix}
u \\
v \\
\end{pmatrix}
=
\frac{1}{\sqrt{\mathrm{Det}(\mathbf{E})}}\, 
\left( \mathbf{E} \cdot \mathbf{B} \right) \,
\begin{pmatrix}
U \\
V \\
\end{pmatrix}
\,,\quad \text{~}\quad
\mathbf{B} = 
\begin{pmatrix}
\frac{1}{\sqrt{2}} & \frac{-\mathrm{i}}{\sqrt{2}}\\
\frac{-\mathrm{i}}{\sqrt{2}} & \frac{1}{\sqrt{2}}\\
\end{pmatrix}~,
\end{equation}
where $\mathbf{E}$ is a 2$\times$2 matrix with columns any two eigenvectors
$e_{1,2}$ associated with the eigenvalues $\lambda_{1,2} = \pm \omega_s$
of the matrix $\mathbf{M}$
\begin{equation}
\mathbf{M} = 
\begin{pmatrix}
\mathtt{a}_{(1,1,0,0)} & 2 \mathtt{a}_{(2,0,0,0)} \\
- 2 \mathtt{a}_{(0,2,0,0)} & -\mathtt{a}_{(1,1,0,0)} \\
\end{pmatrix}~.
\end{equation}
From the variables $U$ and $V$ we then pass to the action-angle variables
$({\cal Y},\phi_f)$ with
\begin{equation}
\begin{aligned}\label{eq:ys_phis}
&U = \sqrt{2Y_s} \sin \phi_s~, &V  = \sqrt{2Y_s} \cos \phi_s~.
\end{aligned}
\end{equation}

The last step corresponds to a re-organization of the terms of the
Hamiltonian, according to a \emph{book-keeping} parameter~\cite{EfthyLaPlata}. 
This is a parameter with numerical value equal to $\epsilon=1$. To every 
term in the Hamiltonian~\eqref{eq:hb_asym_expand}, we asign 
a power of $\epsilon$ indicating the order of the normalization at which 
the term will be treated. Thus, coefficients with powers of $\epsilon$ 
propagate throughout the series at all normalization steps, helping to 
organize the terms in different orders of smallness. Regarding the original 
Hamiltonian, we adopt the following book-keeping rule:
\begin{bookrule}\label{bkp:rule-2}
To every monomial of the type 
\begin{equation*}
\mathtt{c}_{(k_1,k_2,k_3,k_4)}\, (\sqrt{Y_s})^{k_1} (\sqrt{{\cal Y}})^{k_2}\, 
{\textstyle {{\cos}\atop{\sin}}} (k_3 \phi_s + k_4\,\phi_f)~,
\end{equation*}
assign a book-keeping coefficient $\epsilon^{r(k_1,k_2,k_4)}$, where the 
exponent $r(k_1,k_2,k_4)$ is given by
\begin{equation*}
 r(k_1,k_2,k_4) =
   \begin{cases}
   \mathrm{Max}(0\,,\,k_1+k_2-2) & \text{if }\,k_4 = 0 \\
   \mathrm{Max}(0\,,\,k_1+k_2-2)+1 & \text{if }\,k_4 \neq 0 
   \end{cases}~.
\end{equation*}
\end{bookrule}
This book-keeping rule ensures also that the terms of zero-th order in 
$\epsilon$ are linear in ${\cal Y}$ and $Y_s$. The Hamiltonian now takes 
the form:
\begin{equation}
\begin{aligned}\label{eq:hb_asym_expand_lambd_4}
H_b (Y_s,{\cal Y},\phi_s,\phi_f,Y_p) = &\, Y_p  \, + 
\, \omega_s \, Y_s \, +\, 
\omega_f \, {\cal Y} \,\\
+ & \,\sum_{r=1}^{r_{max}}
\mathtt{c}_{(k_1,k_2,k_3,k_4)}\, \epsilon^r (\sqrt{Y_s})^{k_1} 
(\sqrt{{\cal Y}})^{k_2}\, 
{\textstyle {{\cos}\atop{\sin}}} (k_3 \phi_s + k_4\,\phi_f)~.
\end{aligned}
\end{equation}

From the canonical transformation in Eq.~\eqref{eq:ys_phis}, it is
straighforward to check that the harmonics of the angles $\phi_f$ and
$\phi_s$ have the same parity as the powers of the corresponding
functions in the variables $\sqrt{{\cal Y}}$ and $\sqrt{Y_s}$.

\subsection{Resonant normalization}\label{sec:resform}
The resonant normalization of the Hamiltonian~\eqref{eq:hb_asym_expand_lambd_4} 
consists of a sequence of near-identity canonical transformations, in 
ascending powers of the book-keeping parameter $\epsilon$, aiming to 
eliminate from the Hamiltonian the trigonometric dependence on the angles in any 
linear combination other than the one which corresponds to  
the selected secondary resonance (Eq.~\ref{eq:main_sec_res}). The resulting 
normal form includes, besides terms depending just on the actions,
also terms of the form
\begin{equation}\label{eq:struc-res-nf}
b (\mathbf{p}^{(r)}) \,
\mathrm{e}^{\,\mathrm{i}(\mathbf{k} \cdot \mathbf{q}^{(r)})}~.
\end{equation}
Here, $\mathbf{q}^{(0)} = (\phi_f, \phi_s)$, $\mathbf{p}^{(0)}=({\cal
  Y}, Y_s)$, and the superscript $(r)$ indicates the variables found
after $r$ consecutive near-identity normalizing transformations
of $(\mathbf{q}^{(0)},\mathbf{p}^{(0)})$.  Also,
$\mathbf{k}=(k_1,k_2)$ belongs to the set ${\cal M}$ called the
resonant module (Eq.~\ref{eq:res_mod_gen} below).  The
terms~\eqref{eq:struc-res-nf} allow to determine the theoretical
separatrices of the secondary resonance via the process described in
subsection \ref{subsec:sepdef} below.

The general recursive resonant normalization algorithm is defined as
follows: Let $m_1$, $m_2$ be two integers marking the secondary
resonance $\frac{m_1}{m_2} \approx \frac{\omega_f}{\omega_s}$. The
resonant module ${\cal M}$ is the set of integer vectors defined by
\begin{equation}\label{eq:res_mod_gen}
{\cal M} = \{ \mathbf{k} = (k_1,k_2): k_1\, m_1 + k_2\, m_2 = 0 \}~,
\end{equation}
where $\sum_{i=1}^{2} |m_i| \neq 0$. 

Let us assume that the Hamiltonian is in normal form up to order $r$ 
in the book-keeping
parameter, i.e.
\begin{equation}
{\cal H} = {\cal Z}_0 + \epsilon {\cal Z}_1 + \ldots + 
\epsilon^r {\cal Z}_r +
\epsilon^{r+1} {\cal H}^{(r)}_{r+1} + \epsilon^{r+2} {\cal H}^{(r)}_{r+2}
 + \ldots~.
\end{equation}

From the terms of order
$\epsilon^{r+1}$, in the Fourier expansion,
\begin{equation}
{\cal H}^{(r)}_{r+1} = \sum_{\mathbf{k}} b (\mathbf{p}^{(r)}) \,
\mathrm{e}^{\,\mathrm{i}(\mathbf{k} \cdot \mathbf{q}^{(r)})}~,
\end{equation}
where we isolate the terms that we want to
eliminate in the present step, denoted by
\begin{equation}\label{eq:phantomHrp1}
\phantom{{\cal H}}^{\ast}{\cal H}^{(r)}_{r+1} = 
\sum_{\mathbf{k} \notin {\cal M}} b (\mathbf{p}^{(r)}) \,
\mathrm{e}^{\,\mathrm{i}(\mathbf{k}\, \cdot\, \mathbf{q}^{(r)})}~.
\end{equation}
The homological equation
\begin{equation}\label{eq:homol_eq_res_nf}
\epsilon^{r+1}\phantom{|}^{\ast}{\cal H}^{(r)}_{r+1} +
\{ {\cal Z}_0,\chi_{r+1} \} = 0
\end{equation}
has the solution
\begin{equation}\label{eq:gene_func_res_nf}
\chi_{r+1} = \epsilon^{r+1}
\sum_{\mathbf{k} \notin {\cal M}} \frac{ b (\mathbf{p}^{(r)})}
{\mathrm{i\,(\mathbf{k}\, \cdot \,\boldsymbol{\omega})}} \,
\mathrm{e}^{\,\mathrm{i}(\mathbf{k}\, \cdot\, \mathbf{q}^{(r)})}~,
\end{equation} 
with $\boldsymbol{\omega} = (\omega_f,\omega_s)$. 

Having the expression of the generating function, we compute the
transformed Hamiltonian
\begin{equation}\label{hnormrplusone}
{\cal H}^{(r+1)} = \exp({\cal L}_{\chi_{r+1}}) {\cal H} ^{(r)}~,
\end{equation}
where
\begin{equation}\label{eq:Lie-exp-oper}
\exp\Big({\cal L}_{\chi}\Big)\, \cdot \, = \mathbb{I}\,\cdot\, 
 +  ({\cal L}_{\chi} \, \cdot \,) + \frac{1}{2}  
({\cal L}^2_{\chi} \, \cdot \,) + \ldots~.
\end{equation}
and the \emph{Lie operator} ${\cal L}_{\chi} \equiv \{\cdot,\chi\}$ 
($\{\cdot,\cdot \}$ denotes the Poisson bracket).

By construction, the Hamiltonian in Eq.~\eqref{hnormrplusone} is in
normal form up to order $\epsilon^{r+1}$, i.e.
\begin{equation}
{\cal H} = {\cal Z}_0 + \epsilon {\cal Z}_1 + \ldots 
+ \epsilon^r {\cal Z}_r +
\epsilon^{r+1} {\cal Z}_{r+1} + \epsilon^{r+2} {\cal H}^{(r)}_{r+2} + 
\epsilon^{r+3} {\cal H}^{(r)}_{r+3} + \ldots~.
\end{equation}

\subsection{Computation of theoretical separatrices}\label{subsec:sepdef}
Let us consider the function $H_b$ given in
Eq.~\eqref{eq:hb_asym_expand_lambd_4} as the starting Hamiltonian
$H_b^{(0)}$ of the normalizing scheme. We apply the normalizing scheme
presented above, up to a maximum normalization order $R$ in
$\epsilon$.  In the examples that follow, the maximum
  normalization order examined was $R=22$. However, since the resonant
  normal form series are asymptotic, depending on the parameters and
  resonance considered, the optimal normalization order (yielding the
  minimum remainder as computed e.g. in~\cite{Efthy-04}) varies,
  yielding optimal orders between $R=14$ and $R=20$.

Let $H_b^{(R)}$ be 
the final normalized Hamiltonian. According to Eq.~\eqref{eq:struc-res-nf},
the form of $H_b^{(R)}$ is given by
\begin{equation}\label{eq:nf-Hb-con-lambd}
H_b^{(R)} = \sum_{\substack{r=0\\(k_f,k_s) \in {\cal M}}}^{R} 
\epsilon^r \mathtt{b} ({\cal
Y}^{(R)},Y_s^{(R)}) \, \mathrm{e}^{\,\mathrm{i}(k_f \phi_f^{(R)} 
+ k_s \phi_s^{(R)})}~.
\end{equation}
If we replace the book-keeping parameter $\epsilon$ for its value equal 
to 1, we recover the final normal form, depending on the actions and the
angles through the combination,
\begin{equation}\label{eq:nf-Hb-con-lambd}
H_b^{(R)} = \sum_{(k_f,k_s) \in {\cal M}} \mathtt{c}_{(d_f,d_s,k_f,k_s)}\, 
\sqrt{{\cal
Y}^{(R)}}^{\,d_f}\, \sqrt{Y_s^{(R)}}^{\,d_s} \, 
\mathrm{e}^{\,\mathrm{i}(k_f \phi_f^{(R)} + k_s \phi_s^{(R)})}~,
\end{equation}
where the pairs $(d_f,k_f)$ and $(d_s,k_s)$ have the same parity, and
the values of the Fourier wavenumbers are bounded by $|k_f|\leq d_f$ and
$|k_s|\leq d_s$.  The integers $(d_f,d_s)$ are limited by the value of
$R$, through the book-keeping Rule~\ref{bkp:rule-2}. 

We define the quantity 
\begin{equation}\label{eq:resint}
\Psi = m_1\,{\cal Y}^{(R)} + m_2 \, Y_s^{(R)}
\end{equation}
as a \emph{resonant integral} of the normal form $H_b^{(R)}$, 
where $m_1$ and $m_2$ are the integers that define the
resonant module ${\cal M}$ in Eq.~\eqref{eq:res_mod_gen}.
Considering Eq.~\eqref{eq:struc-res-nf}, it is straightforward to prove
that
\begin{equation}\label{eq:res-intgral-cond}
{\cal L}_{H_b^{(R)}}\, \Psi = \{H_b^{(R)}, \Psi\} = 0~,
\end{equation}
i.e. $\Psi$ is a formal integral of $H_b^{(R)}$. 

By considering the transformation $\mathscr{C}^{(R)}$,
\begin{equation}\label{eq:transfC}
\mathscr{C}^{(R)} = \varphi^{(1)} \circ \varphi^{(2)} \circ \ldots \circ
\varphi^{(R-1)} \circ \varphi^{(R)}~,
\end{equation}
where
\begin{equation}\label{eq:varphi-r}
\varphi^{(r)} = \exp\left({\cal L}_{\chi_r} \right)({\cal Y}^{(r)},Y_s^{(r)},
\phi_f^{(r)}, \phi_s^{(r)})
\end{equation}
we can represent the resonant integral in terms of the original
variables $({\cal Y}^{(0)},Y_s^{(0)},\phi_f^{(0)},\phi_s^{(0)})$, via
\begin{equation}\label{eq:non-norm-res-int}
\Psi({\cal Y}^{(0)},Y_s^{(0)},\phi_f^{(0)},\phi_s^{(0)}) = 
\Psi\left(\mathscr{C}^{(R)}({\cal Y}^{(R)},Y_s^{(R)},\phi_f^{(R)},
\phi_s^{(R)}) \right)~.
\end{equation}
Finally, applying the inverse transformations to those of 
Eqs.~\eqref{eq:ys_phis} and~\eqref{eq:fromuvtoUV}, we are able to
express the resonant integral in \eqref{eq:non-norm-res-int} as
function of the variables used in Eq.~\eqref{eq:hb_asym_expand}
\begin{equation}\label{eq:uvYphif-res-int}
\Psi \equiv \Psi(v,{\cal Y},u,\phi_f)~.
\end{equation}

Having arrived at a final expression for the resonant integral $\Psi$ 
in terms of the original canonical variables, we can compute the form 
of the theoretical separatrices of the corresponding secondary resonance 
in any suitably defined surface of section of the Hamiltonian $H_b$. 
In the numerical results below, we adopt
a section of the form $\phi_f=\phi_{f0}$, as well as a 
constant value of the energy $E=H_b$, the equation
$E=H_b(v,{\cal Y},u,\phi_{f0})$ can be solved for ${\cal Y}$. 
Substitution to~(\ref{eq:uvYphif-res-int}) yields then the resonant 
integral on the surface of section as a function of $u$ and $v$ 
only, viz.
\begin{equation}\label{eq:uvYphif-res-int-sos}
\Psi \equiv \Psi \Big( v,{\cal Y}(u,v;E,\phi_{f0}),u,\phi_{f0} \Big)~.
\end{equation}
The theoretical phase portrait is now obtained by the level curves 
of Eq.~\eqref{eq:uvYphif-res-int-sos}. Figure \ref{fig:minresint.jpg},  
left panel, summarizes the main features of the theoretical phase 
portrait. In particular, the stable periodic orbit of the secondary 
resonance is represented by the points of extremum of the level set 
of $\Psi$, while the unstable periodic orbit corresponds to the minimax 
(saddle) points of the level set of $\Psi$. The level curves with 
$\Psi=\Psi_{nmx}$, where $\Psi_{nmx}$ is the value of the resonant 
integral at the saddle points, are the curves representing the 
theoretical separatrices of the secondary resonance.  

\section{Numerical results: boundary of the effective 
stability domain}

\subsection{Analytical vs. numerical stability boundary}
\begin{figure}[t]
\centering
\includegraphics[width=0.99\textwidth]{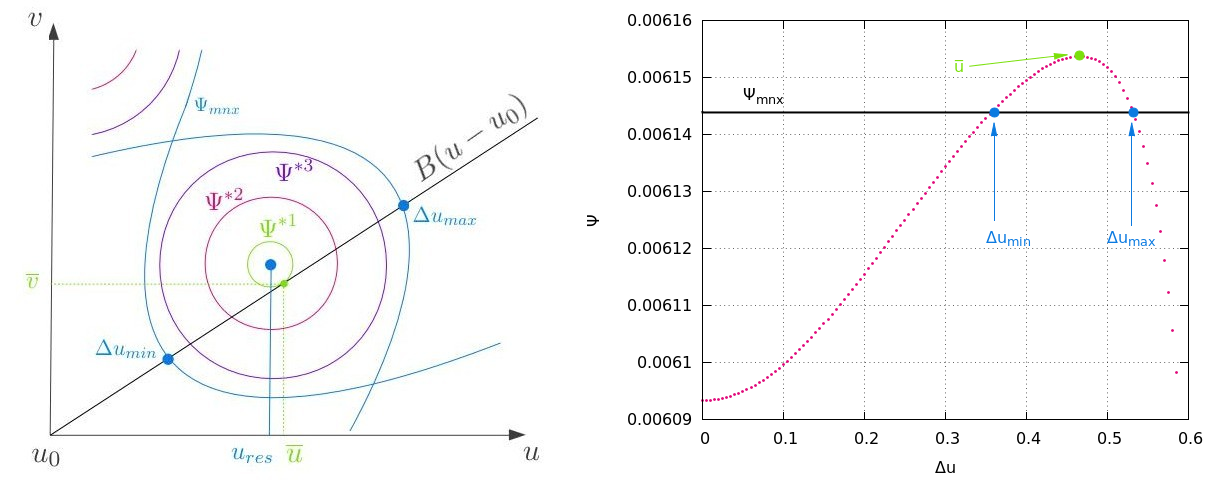}
\caption[Schematic representation of the level curves of the resonant
  integral in the plane $(u,v)$. The resonant integral as function of
  $\Delta u$]{Left panel - Schematic representation of the plane
  ($u,\,v$) for a surface of section of the $H_{b}$. The central blue
  dot represents the location of a stable periodic orbit, whose
  co-ordinate in equal to $u=u_{res}$. At this point, the resonant
  integral $\Psi$ presents a global extremum. Additional
  quasi-periodic orbits inside the island of stability are labeled
  with the corresponding values of $\Psi$, i.e. $\Psi^{\ast 1}$,
  $\Psi^{\ast 2}$, $\Psi^{\ast 3}$, $\Psi_{mnx}$, accomplishing
  $\Psi^{\ast 1} > \Psi^{\ast 2} > \Psi^{\ast 3} > \Psi_{mnx}$. The
  value $\Psi_{mnx}$ represents a theoretical separatrix of the
  resonance in the resonant integral approximation (in reality,
  instead of the separatrix we have a thin separatrix-like chaotic
  layer). For the initial conditions taken along the line $B(u-u_0)$,
  the orbit for which $\Psi$ is maximum corresponds to a level
  curve tangent to the line, labeled $\Psi^{\ast 1}$. The initial
  condition for u along this line, $\overline{u}$, represents a good
  approximation to the exact resonant position $u_{res}$. The two
  values of $u$ on the line $B(u-u_0)$ satisfying $\Psi = \Psi_{mnx}$
  correspond to the intersection of the separatrix with the line
  $B(u-u_0)$ ($\Delta u_{min}$ and $\Delta u_{min}$, in blue), and
  provide an estimation of the width of the resonance. Right panel -
  Values of the resonant integral $\Psi$ along the line
  $B(u-u_0)$. The position of the maximum of the function corresponds
  to $\overline{u}$ (green dot). The value of $\Psi_{mnx}$ (black
  line) defines the position of the two borders of the resonance
  $\Delta u_{min}$ and $\Delta u_{max}$ (blue dots).}
\label{fig:minresint.jpg}
\end{figure}

We present below numerical results based on the computation of
stability maps for selected values of the parameters $\mu$ and $e'$,
characterized by the presence of conspicuous secondary resonances of
the Hamiltonian $H_b$. The stability maps are given in color scale of
the values of the Fast Lyapunov Indicator~(\cite{FLI-00}), for orbits
with initial conditions labeled in terms of two quantities $(\Delta
u,e_{p_0})$. These quantities also serve as proper elements,
i.e. quasi integrals of motion, for the subset of all regular orbits
in every stability map. Working on fixed surfaces of section $\phi_f =
-\pi/3$, the relation between initial conditions $(u,v,{\cal Y})$ and
$(\Delta u,e_{p_0})$ is given by the relations ${\cal Y}=
\frac{e_{p,0}^2}{2}$, $\Delta u=u-u_0$, where $u_0$ is the point of
intersection of the short-period orbit around L4 with the surface of
section (see~\cite{PaezEfthy2015} for analytical expressions), and $v
= B\Delta u$, for fixed parameters $B$ (depending on $\mu$) selected
in such a way that the straight line $v=B(u-u_0)$ in the surface of
section passes right through one of the islands of the secondary
resonance chain. The half-witdh of the libration in $u$ as a
  function of $\Delta u$, $B$, $\mu$, $e_p$ and $e'$, reads
\begin{equation}\label{dudp}
D_p =  \left[ \frac{3B^2/2 + \mu \left( 9/8 + 63e'^2/16 
+ 129 e_p^2/64 \right)}{\mu \left( 9/8 + 63 e'^2/16 
+ 129 e_p^2/64 \right)}  \right]^{1/2} \Delta u 
+ {\cal O}(\Delta u^2)~.
\end{equation}
The values of $B$ used in the various stability maps below are
given explicitly in the caption of each figure.

We can now
superpose the theoretical computation of the phase portrait of the
secondary resonance to the numerical results found in the stability
maps. For given parameters $\mu,e'$, $B$, 
 and choosing one value of the energy $E$, one
obtains the resonant integral \eqref{eq:uvYphif-res-int-sos} as a
function of $u$ only. An example is shown in the right panel of
Fig.~\ref{fig:minresint.jpg}. The value $u=\overline{u}$ marks the
position of local maximum of the resonant integral $\Psi$ along the
line $v=B(u-u_0)$. This corresponds a central locus passing
approximately through the middle of the resonant domain along the
corresponding secondary resonance. On the other hand, the points of
intersection of the line $\Psi = \Psi_{mnx}$ with the curve of the
resonant integral mark the values $u_1,u_2$, and hence $\Delta
u_{min}=u_1-u_0$, $\Delta u_{max} = u_2-u_0$, where the theoretical
separatrix intersects the plane of the stability map.  The
corresponding values of $e_{p0}$ can be found through $e_{p0,i} =
\Big[ -2{\cal Y} \Big(u_i,v_i=B(u_i-u_0),\phi_{f0}\,;\,E \Big)
  \Big]^{1/2}$, with $i=1,2$.

\begin{figure}[t]
\centering
\includegraphics[width=0.70\textwidth]{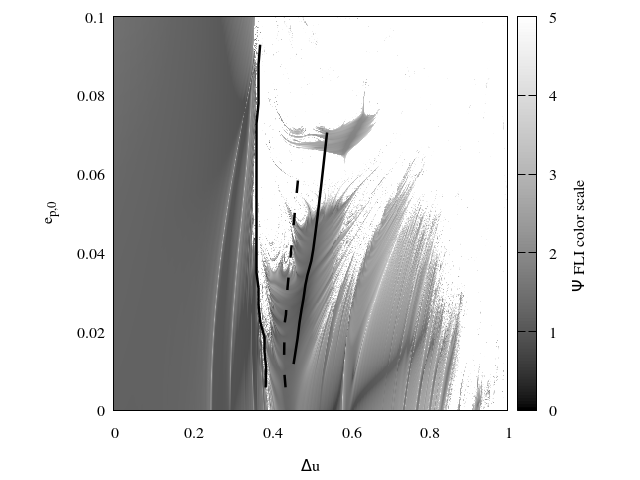}
\caption{Theoretical location of the center and borders of the 1:6
  secondary resonance for $\mu= 0.0041$, $B = 0$ and $e' = 0.02$. 
    The solid lines correspond to the inner an outer border of the
    resonance, the dashed line correspond to the estimation of the
    center of the resonance. The underlying image gives the numerical
    stability map, using the FLI value in grayscale.}
\label{fig:threeelin.png} 
\end{figure}
Repeating, now, the same process for different values of the energy $E$ 
allows to obtain the whole locus of the theoretical center as well as 
the theoretical boundary of the secondary resonance on the FLI stability 
map. Figure~\ref{fig:threeelin.png} shows an example of the 
location of the center and borders of a secondary resonance, with
the method of the resonant integral, for the case of the 1:6 secondary
resonance ($\mu = 0.0041$) and $e' = 0.02$. The position of the center
of the resonance is denoted by a dashed line, and the inner and 
outer borders are denoted by thick solid lines. 
By comparison with the underlaying FLI stability map, we can see that
both the center of the resonance and the outer border $\Delta u_{max}$
are understimated by this computation, proving that the overall
estimation of the resonance width is not accurate. On the other hand,
the key remark is that the method turns to be extremely efficient in 
the location of the inner border. The approximate position of 
$\Delta u_{min}$ is well determined in the whole range of proper 
eccentricity values considered $0 < e_{p,0} < 0.1$.

\begin{figure}[t]
\centering
\includegraphics[width=0.99\textwidth]{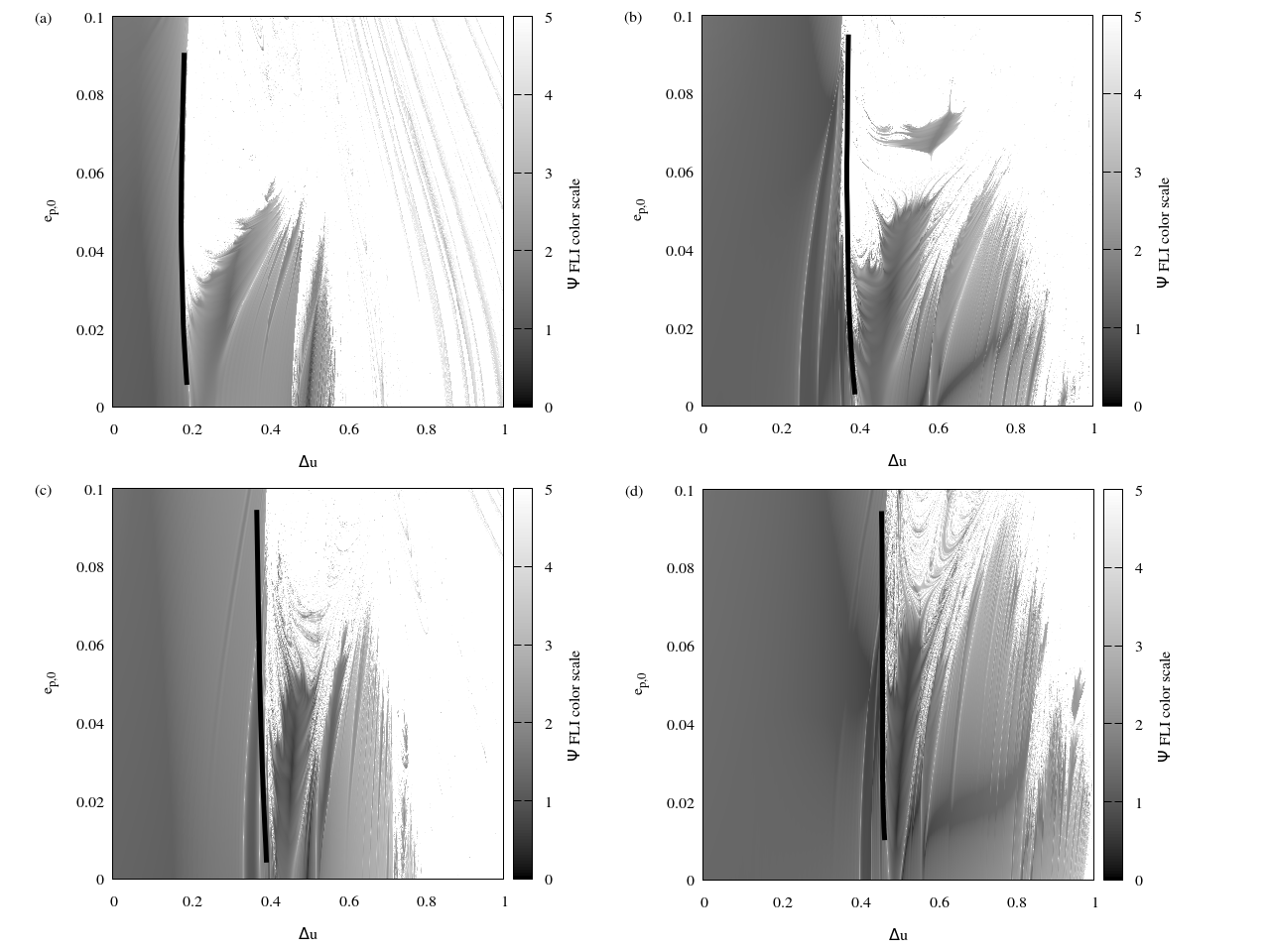}
\caption{Determination of the effective stability domain ($\Delta
  u_{min}$, thick black line), for the secondary resonances $1$:$5$
  ($\mu=0.0056$, $B=0.03$, panel a), $1$:$6$ ($\mu=0.0041$, $B = 0$,
  panel b), $1$:$7$ ($\mu=0.0031$, $B=0.015$, panel c) and $1$:$8$
  ($\mu=0.0024$, $B = 0$, panel d), and $e'=0.02$.}
\label{fig:theflis.png}
\end{figure}

Figure~\ref{fig:theflis.png} shows more examples of the method 
of determination of the effective stability domain through the 
application of the resonant normal form in the cases of the secondary 
resonances $1$:$5$ ($\mu=0.0056$, panel a), $1$:$6$ ($\mu=0.0041$, 
panel b), $1$:$7$ ($\mu=0.0031$, panel b) and $1$:$8$ ($\mu=0.0024$, 
panel d), and primary's eccentricity $e'=0.02$. In all the panels, 
the location of the inner border $\Delta u_{min}$ is shown with a 
thick black line on top of the corresponding FLI stability map. 
We observe that this limit 
divides the space of proper elements in two regions: the inner domain 
from $\Delta u = 0$ to $\Delta u_{min}$ is populated mainly by regular 
orbits, and exhibits also some isolated resonances of small width, in 
which the orbits can only be weakly chaotic and remain practically stable. 
On the contrary, the domain external to $\Delta u_{min}$ is dominated 
by the presence of conspicuous resonances as well as regions of 
strong chaos. It is remarkable that the analytical determination of 
the inner border of the resonances, which is based on an integrable 
approximation to the Hamiltonian (i.e. the resonant normal form), 
can still provide an accurate limit even in domains of the phase 
space where the resonant orbits are, in reality, chaotic. It is 
this robustness of the inner border determination which renders 
the whole approach useful in practice. 

\begin{figure}[t]
\centering
\includegraphics[width=0.99\textwidth]{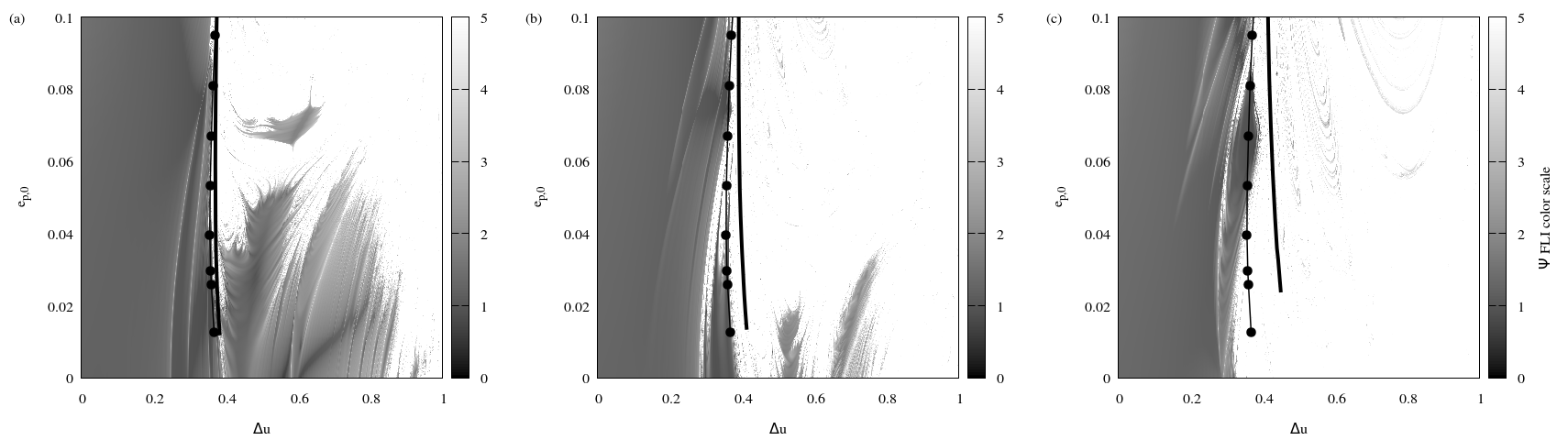}
\caption{FLI stability maps for the $1$:$6$ secondary resonance ($\mu
  = 0.0041$, $B=0$) and three values of the eccentricity $e'=0.02$
  (a), $e'=0.06$ (b), $e'=0.1$ (c). The dotted thin line
    corresponds to the analytical determination of $\Delta u_{min}$
    for the parameters $\mu = 0.0041$ and $e'= 0$ (circular case) in
    all three panels, while the thick line corresponds to $e'=0.02$ in
    (a), $e'=0.06$ in (b) and $e'=0.1$ in (c). }
\label{fig:1to6panels.png}
\end{figure}

\begin{figure}[t]
\centering
\includegraphics[width=0.99\textwidth]{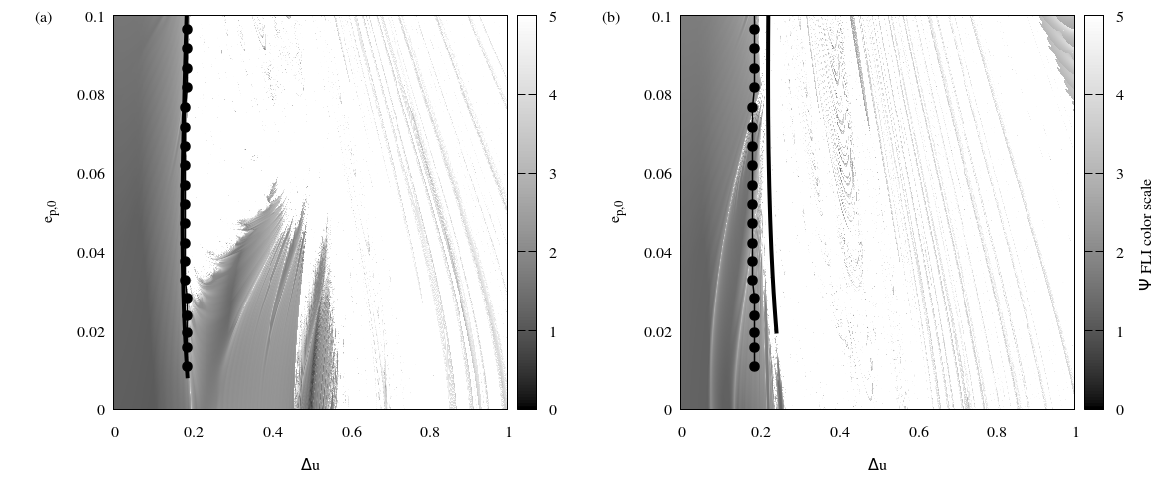}
\caption{FLI stability maps for the $1$:$5$ secondary resonance ($\mu
  = 0.0056$, $B=0.03$) and two values of the eccentricity $e'=0.02$
  (a), $e'=0.08$ (b).  The dotted thin line corresponds to the
    analytical determination of $\Delta u_{min}$ for the parameters
    $\mu = 0.0056$ and $e'= 0$ (circular case) in the two panels,
    while the thick line corresponds to $e'=0.02$ in (a) and $e'=0.08$
    in (b). }
\label{fig:1to5panels.png}
\end{figure}

Figures~\ref{fig:1to6panels.png} and~\ref{fig:1to5panels.png} show,
now, more examples of the applicability as well as the level of
approximation of the method. Figure~\ref{fig:1to6panels.png} shows the
stability maps for $\mu=0.0041$
(corresponding to a conspicuous $1$:$6$ secondary resonance) and three
different values of the primary's eccentricity, $e'=0.02$ (panel a),
$e'=0.06$ (panel b), $e'=0.1$ (panel c).  In the same plots we show
the effective stability borders from the resonant normal form
computation for the 1:6 secondary resonance, but for two values of the
primary's eccentricity in each case, namely $e'=0$ (dotted thin
  line) and $e'=0.02$ (thick line) in (a), $e'=0$ (dotted thin line)
  and $e'=0.06$ (thick line) in (b) and $e'=0$ (dotted thin line) and
  $e'=0.1$ (thick line) in (c).  We observe that altering the
primary's eccentricity from $e'=0$ to only $e'=0.1$ suffices to
completely wipe out the entire structure of secondary resonances
beyond $\Delta u\simeq 0.4$. In fact, we observe that, with increasing
$e'$, so called `transverse' resonances, i.e.  involving also the
secular frequency $g$, i.e. of the form $m_f \omega_f + m_s \omega_s +
m_g g = 0$ with $m_g \neq 0$, appear near this border. For example,
the $1$:$6$:$1$ resonance at $u \sim 0.25$ in panel (a) of
Fig.~\ref{fig:1to6panels.png} moves towards the border at $u\approx
0.35$ in panel C of the same figure. A careful inspection of the
stability maps shows that, for small $e'$ these resonances have a
small width and remain isolated within the inner stability domain,
while, as $e'$ increases, all resonances (main or transverse) grow in
size and move outwards, until they enter to the region of strong
chaos. As revealed in the panels of Fig.~\ref{fig:1to6panels.png},
these two effects (the moving of the resonances outwards and the
refilling of the stable region with transverse resonances) counteract
each other in such a way that the border separating the inner domain
of stability from the outer chaotic domain remains practically in the
same place.  Due to this effect, we can see that even the estimation
of the border via the resonant normal form corresponding to the {\it
  circular} case ($e'=0$, dotted thin line) practically suffices to
obtain a good approximation of the border of the effective stability
domain. Also, regarding the Trojan's body eccentricity,
  parameterized by $e_{p,0}$, one remarks that stable domains of all
  the secondary resonances, beyond the main stability domain, survive
  only for small values of $e_{p,0}$.  This is because the amplitude
  of the separatrix pulsation increases as the eccentricity of the
  Trojan body increases. As a consequence, we find that the border of
  the main domain of stability is more sharp, and, thus, in general,
  better represented by the analytical resonance limit as $e_{p,0}$
  increases.

Similar results are found in Fig.~\ref{fig:1to5panels.png}, showing
the stability maps for $\mu = 0.0056$, corresponding to a conspicuous
$1$:$5$ secondary resonance, and for the primary's eccentricity values
$e'=0.02$ (panel a), $e'=0.08$ (panel b). The estimated borders are
found by the resonant normal form determination for $\mu=0.0056$,
using the parameters $e'=0$ (circular case, dotted thin line) and
$e'=0.02$ (thick line) in (a), and $e'=0$ (dotted thin line) and $e=
0.08$ (thick line) in (b).  The margin between the two theoretical
curves is again small (of about $0.02$ rad in $\Delta u$), while,
again, the determination of the border of the stability domain using
the circular model suffices to practically obtain an accurate limit of
the domain of stability.  In fact, in both Figures
  \ref{fig:1to6panels.png} and \ref{fig:1to5panels.png} the extent
  occupied by the stable parts of the corresponding resonances is
  determined by the separatrix pulsation effect. The amplitude of
  the pulsation depends on terms absent from the `basic model', thus
  this effect cannot be modelled using only the resonant integrals of
  the basic model.  However, as a rule of thumb we find that the
  border of the domain of stability lies always between two
  theoretical border determinations by the basic model, i.e., one
  using the circular model $e'=0$ and a second using a moderate value
  of the primary's eccentricity, e.g. $e'=0.1$.

\subsection{Robustness with respect to parameter values}
The investigation in the previous subsection focused on particular
values of $\mu$ selected with the criterion that, for low
eccentricities of either the primary perturber or the test body
($e',e_{p,0}<0.1$), the phase space of the basic model is dominated by
a low-order secondary resonance of the form 1:$n$ with
$n=5,6,...$. Repeating a comparison between FLI maps and innermost
separatrix borders of secondary resonances, a behavior similar to
Figures \ref{fig:1to5panels.png} (resonance 1:5, for $\mu=0.0056$)
and \ref{fig:1to6panels.png} (resonance 1:6, for $\mu=0.0041$) for
low eccentricities is found when one considers the resonances 1:7
for $\mu=0.0031$, 1:8 for $\mu=0.0024$, 1:9 for $\mu=0.0021$,
1:10 for $\mu=0.0016$, 1:11 for $\mu=0.0014$, 1:12 for
$\mu=0.0012$. These values of $\mu$ are shifted positively with
respect to the bifurcation values $\mu=\mu_{1:n}$ of each
corresponding 1:$n$ short period family in the basic model.  The
shift reflects the fact that, keeping $e',e_{p,0}$ constant, and
increasing $\mu$ as $\mu = \mu_{1:n}+\Delta\mu$, with $\Delta\mu>0$,
the resonance 1:$n$ moves outwards from the libration center, i.e.,
towards higher libration amplitudes $\Delta u$, as $\Delta\mu$
increases. In the resonant integral approximation, the outward motion
of each resonance is accompanied by an increase of its separatrix
width. However, the integrable aproximation fails due to resonance
overlap with nearby resonances as $\Delta\mu$ increases. This
antagonism between outward expansion and resonance overlap determines
the real limit of the domain of stability (see~\cite{Voglis-98},
\cite{Efthyelat-99} for a description of this phenomenon in simple
dynamical maps).

The bifurcation value $\mu_{m:n}$ for the $m$:$n$ short-period family 
of the basic model can be estimated by the root for $\mu$ of the 
equation:
\begin{equation}\label{eq:mubif}
m(1-27\mu/8)=
n\sqrt{6\mu\left({9\over 8}+{63e'^2\over 16}+{129e_p^2\over 64}\right)} 
\end{equation}

Applying Eq.~(\ref{eq:mubif}) to the 1:6 resonance, we find
$\mu_{1:6}\approx 0.0040$ for $e'=e_{p}=0.02$, while
$\mu_{1:6}\approx 0.0038$ for $e'=e_{p}=0.1$. As evident from
Fig.~\ref{fig:1to6panels.png}, the resonance is clearly dominant at
$\mu=0.0042$. In fact, we find that the $1$:$6$ resonant integral
inner separatrix limit applies already when $\Delta \mu\geq 0.001$
with respect to the bifuration value for low eccentricities. On the
other hand, as shown in panels (a) and (e) of Fig.~\ref{fig:from6to5.png},
the analytical series computation with the 1:6 resonant integral
starts collapsing when $\mu=0.0044$, or $\Delta\mu\approx 0.004$. In
practice, the whole separatrix domain around the 1:6 resonance has
been transformed into a chaotic domain. Thus, while it remains true
that the 1:6 resonance of the basic model delimits the main
stability domain, the convergence of the series representing the
theoretical computation of the corresponding resonant integral becomes
poor.

\begin{figure}[t]
\centering
\includegraphics[width=0.99\textwidth]{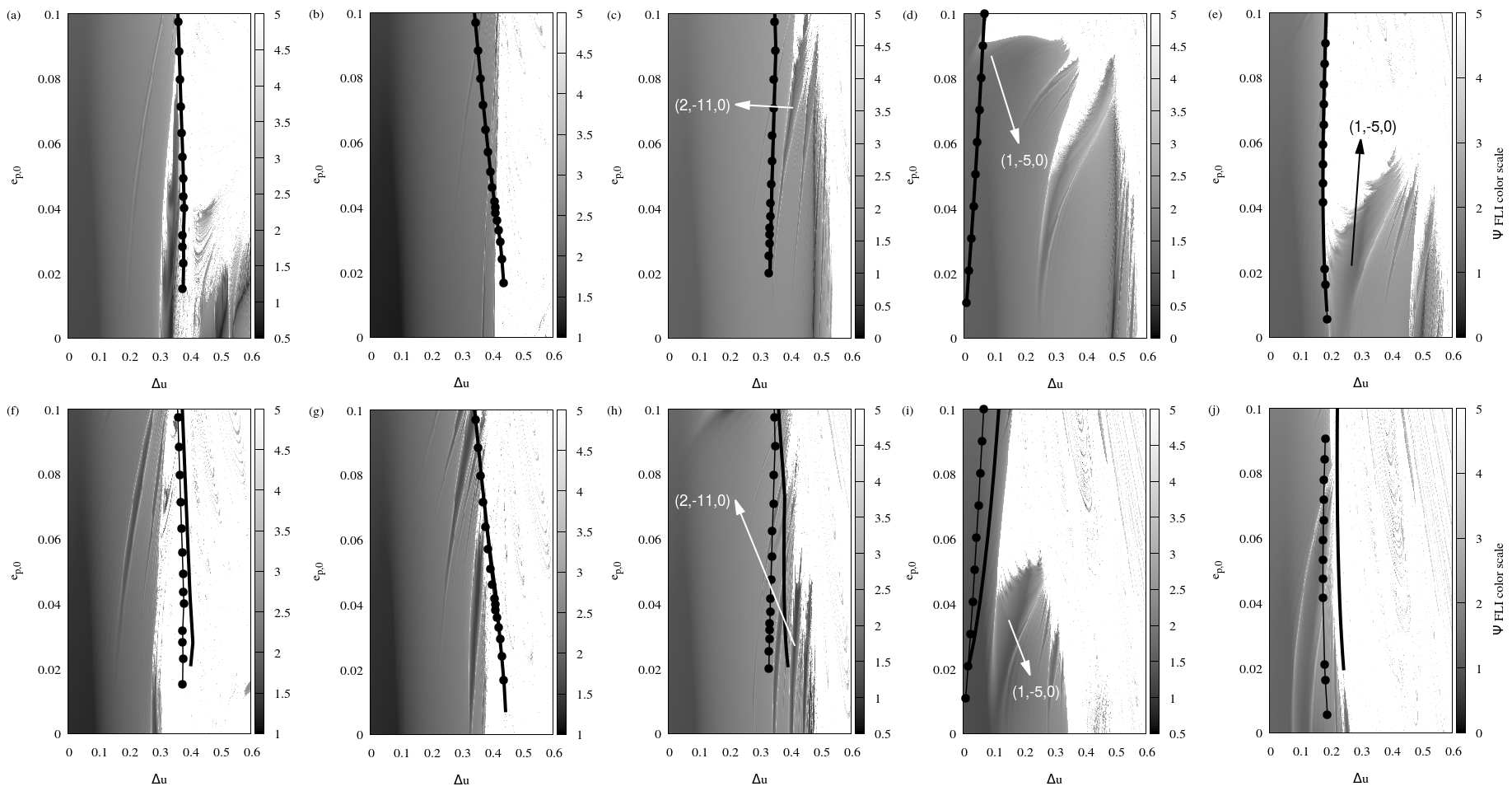}
\caption{FLI stability maps for $\mu=0.0042$ and $e'=0.02$ (a),
  $\mu=0.0044$ and $e'=0.02$ (b), $\mu=0.0048$ and $e'=0.02$ (c),
  $\mu=0.0054$ and $e'=0.02$ (d), $\mu=0.0056$ and $e'=0.02$ (e),
  $\mu=0.0042$ and $e'=0.08$ (f), $\mu=0.0044$ and $e'=0.08$ (g),
  $\mu=0.0048$ and $e'=0.08$ (h), $\mu=0.0054$ and $e'=0.08$ (i), and
  $\mu=0.0056$ and $e'=0.08$ (j).  The thick lines yield the analytical
  estimation of the border of stability for the corresponding value of
  $\mu$ and $e'=0.02$ for the upper row panels, and $e'=0.08$ for the
  lower row panels. The dotted thin line yields the analytical
  estimation of the border for the corresponding value of
  $\mu$ and $e'=0$ (circular case) in all the panels. $B = 0.03$ for
  all the panels.}
\label{fig:from6to5.png}
\end{figure}
Implementing, now, Eq.~(\ref{eq:mubif}) to the $1$:$5$ resonance we 
find $\mu_{1:5}\approx 0.0057$ for $e'=e_{p,0}=0.02$, while $\mu_{1:5}\approx 
0.0054$ for $e'=e_{p,0}=0.1$. Thus $\mu_{1:5}-\mu_{1:6}\approx 0.0016$, 
which implies that the distance in $\mu$ separating the resonances 
1:6 and 1:5 is about 3-4 times larger than the interval of 
values of $\Delta \mu$ for which the validity of the resonant integral 
computation using one particular resonance is satisfactory. In principle, 
in order to bridge the gap between the two resonances, one has to use 
higher order resonances of the basic model, since the border of the 
domain of stability is always delimited by one such resonance. In 
practice, we find that it suffices to consider the basic resonances 
1:$n$ and their first Farey tree combination, i.e., the resonances 
2:$(2n-1)$ which bifurcate at intermediate values of $\mu$, i.e. 
$\mu_{1:n}<\mu_{2:2n-1}<\mu_{1:n-1}$ 
for fixed $e',e_p$. Figure~\ref{fig:from6to5.png} exemplifies the 
transition from the dominance of the 1:6 to the 1:5 resonance 
via the 2:11 resonance of the basic model, for two values of the 
primary's eccentricity $e'=0.02$ (upper row) and $e'=0.08$ (lower row). 
The collapse of the inner border calculation for the 1:6 resonances 
starts near $\mu=0.0044$. However, 
the computation using the 2:11 resonant integral restores a correct 
estimate of the main domain of stability for $\mu=0.0048$, leaving 
only secondary resonances outside this domain. The 2:11 resonance 
remains dominant in this respect up to $\approx\mu=0.0054$. At this 
value of $\mu$ the 1:5 secondary resonance of the basic model 
bifurcates for large enough values of the eccentricities, a fact 
which implies that the whole domain beyond the innermost separatrix 
of the 1:5 resonance should now be considered as outside the main 
stability domain. Indeed, although these secondary resonances are still 
very stable for very low eccentricities, we see that they essentially 
disappear for values of the eccentricities near $\approx 0.1$ (compare 
panels (d) and (i) of Fig.~\ref{fig:from6to5.png}). This marks the transition 
from the dominance of the 2:11 to the 1:5 resonance, the latter 
one being clearly dominant for a somewhat still higher value of $\mu$ 
($\mu = 0.0056$ in panels e, j). 

\begin{figure}[t]
\centering
\includegraphics[width=0.99\textwidth]{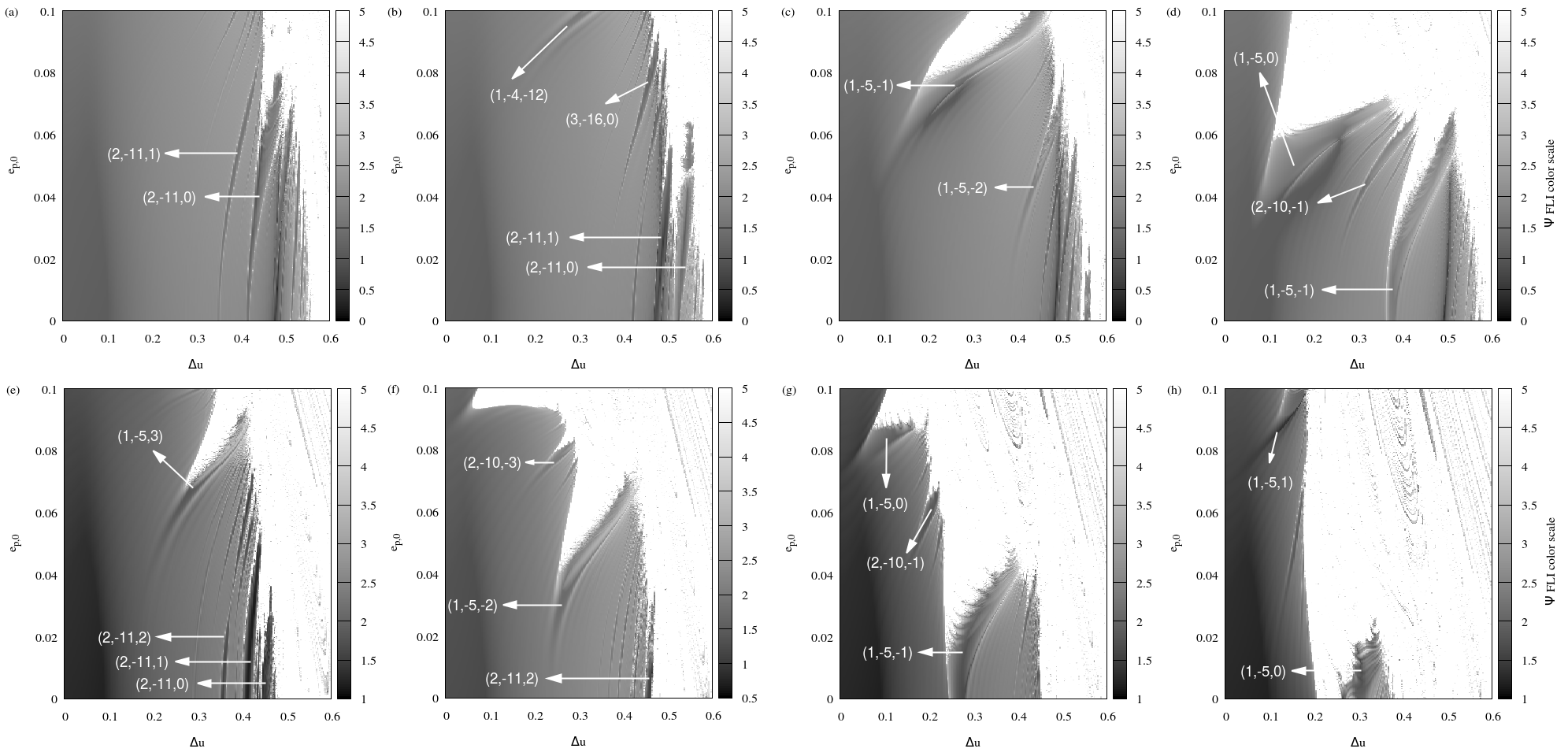}
\caption{Details of the transition from the dominance of the $2$:$11$ 
to the $1$:$5$ resonance as $\mu$ varies from $\mu = 0.0049$ to $\mu=0.0055$.
The parameters in each panel are (a). $\mu=0.0049$, $e'=0.02$, (b). $\mu=0.0051$, 
$e'=0.02$, (c). $\mu=0.0053$, $e'=0.02$, (d). $\mu=0.0055$, $e'=0.02$, 
(e). $\mu=0.0049$, $e'=0.08$, (f). $\mu=0.0051$, $e'=0.08$, (g). $\mu=0.0053$, 
$e'=0.08$, H. $\mu=0.0055$, $e'=0.08$. The domains of various resonances 
(including transverse ones) are marked in the same plots.}
\label{fig:retransition.png}
\end{figure}
Figure \ref{fig:retransition.png} shows in greater detail the transition 
from the $2$:$11$ to the $1$:$5$ resonance, which, using FLI stability 
maps of the full problem, is actually seen to involve also some 
resonances coined {\it transverse} in~\cite{PaezEfthy2015}, i.e. 
resonances involving all three short, synodic and secular frequencies. 
In particular, we see that the border of stability, which for 
$\mu=0.0049$ is practically delimited by the 2:11 resonance, 
starts being gradually penetrated by the transverse resonances 
2:11:1, 1:5:2 and 1:5:1. The penetration appears
earlier, as $\mu$ increases, for higher values of the eccentricities. 
This effect leaves small windows of values of $\mu$ for which, 
for low eccentricities, the border of stability may appear dominated 
by the innermost separatrix of some transverse resonance (e.g. the 
resonance 1:5:1 in panel (c) for $\mu=0.0053$, $e'=0.02$). 
However, for the same value of $\mu$, the innermost separatrix border 
of the $1$:$5$ resonance appears also in the upper part of the stability 
map for higher primary's eccentricity, i.e., $e'=0.08$ (panel g). 
As a consequence, although a clear dominance of the 1:5 resonance 
occurs for all eccentricities beyond $\mu=0.0055$ (panels d, h), 
the 1:5 resonance practically dominates in a wide range of 
eccentricities already at $\mu=0.0053$. In fact, this dominance 
can only become more pronounced when additional perturbations are 
added to the model. 

In conclusion, except for small transient windows of parameter values, 
one can practically always find a resonance of the basic model for which 
the innermost separatrix provides the limit of the main domain 
of stability. It is to be stressed that this is a physical property  
induced by resonant dynamics, which holds independently of the efficiency 
by which the innermost separatrix border of the resonance can be computed 
analytically using some form of resonant integral series. On the other hand, 
using the method presented in Section~\ref{sec:resform}, we find precise 
results by limiting the choice of resonance of the basic model among 
the set 1:$n$ or 2:$(2n-1)$, with $n$ integer. As a rule of thumb, 
for given parameters $\mu,e'$ we choose the limiting resonance as the 
rational number closer to the frequency ratio:
\begin{equation}\label{fratio}
f = 
{\sqrt{6\mu\left({9\over 8}+{63e'^2\over 16}+{129e_p^2\over 64}\right)}
\over
(1-27\mu/8)}
\end{equation}
for values of $e_p$ within our domain of interest. 

\begin{figure}[t]
\centering
\includegraphics[width=0.99\textwidth]{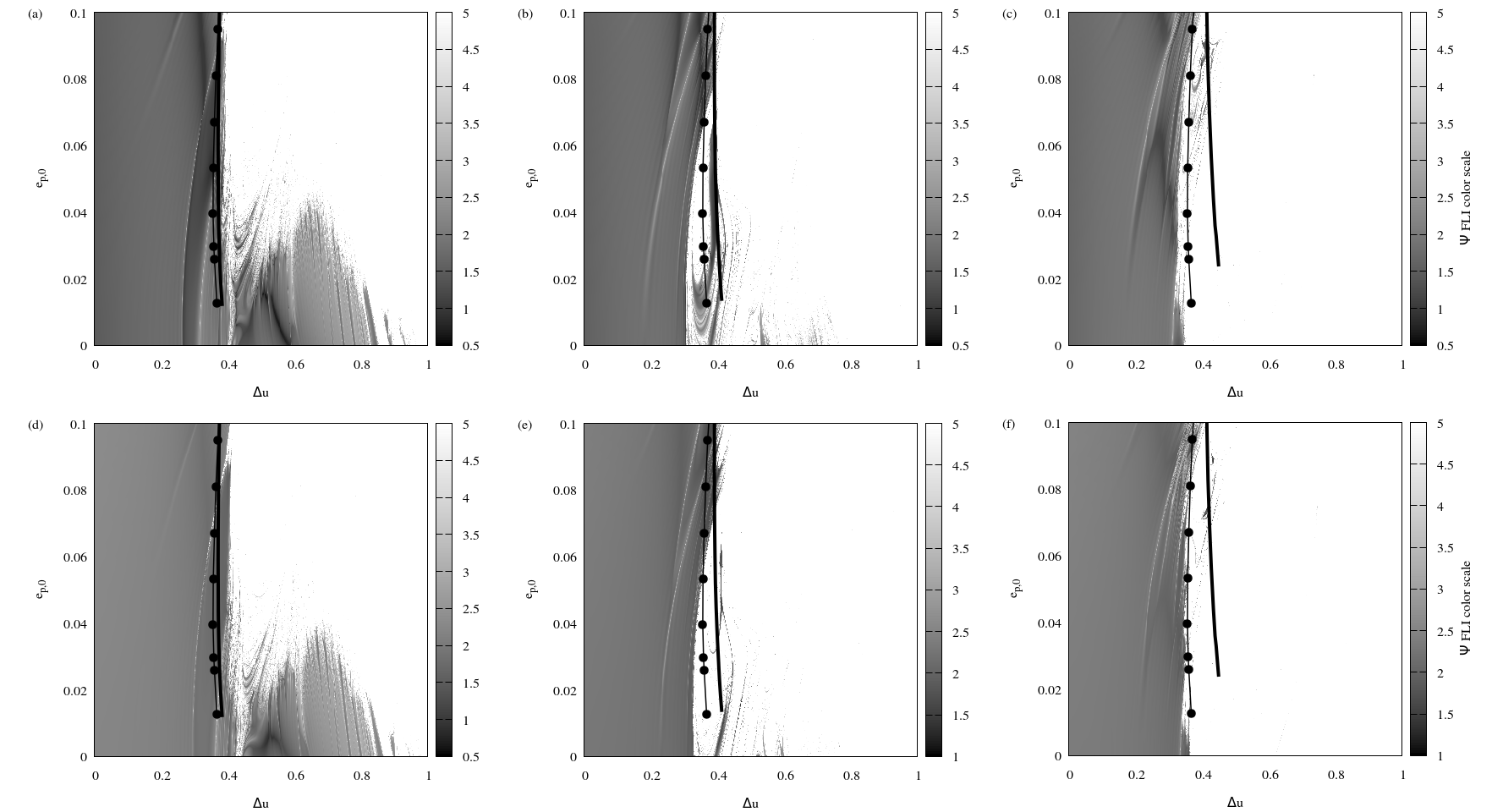}
\caption{The FLI stability maps obtained with the same initial conditions 
as in Fig.\ref{fig:1to6panels.png}, but running the full planar three 
body model instead of the ERTBP, with a mass $\mu_2\neq 0$ assigned to 
the Trojan body. The mass of the primary is always $\mu=0.0041$, while 
the remaining parameters (initial eccentricity $e'$ of the primary and 
mass $\mu_2$ of the Trojan) are (a). $e'=0.02$, $\mu_2=3\times 10^{-6}$, 
(b). $e'=0.06$, $\mu_2=3\times 10^{-6}$, (c). $e'=0.1$, $\mu_2=3\times 10^{-6}$, 
(d). $e'=0.02$, $\mu_2=3\times 10^{-5}$, (e). $e'=0.06$, $\mu_2=3\times 10^{-5}$, 
(f). $e'=0.1$, $\mu_2=3\times 10^{-5}$. The analytical curves are those of
Fig.~\ref{fig:1to6panels.png}.}
\label{fig:full3BP-1to6.png}
\end{figure}

\subsection{Robustness with respect to the choice of model}
As an additional test, we examine the robustness of the above results against
changing the dynamical model for Trojan orbits. Several formation
scenaria discussed in literature~(\cite{Beugetal-07},
\cite{CressNel-09}, \cite{Giuppo-12}, \cite{Lyra-09},
\cite{Pierens-14}) have allowed relatively massive Trojan planets (of
mass $\sim 1$ Earth mass) to exist. Allowing the Trojan body to
have considerable mass, we examine whether the stability borders found
in the framework of the `basic model', which is only derived from the
ERTBP, are still applicable in the framework of the full planar three
body problem.

As an example, Figure \ref{fig:full3BP-1to6.png} compares the border
of stability in the planar ERTBP for $\mu=0.0041$ with one computed in
the full three body problem with the Trojan body having mass equal to
1 or 10 Earth masses.  We consider the Hamiltonian in Poincar\'{e}
variables:
\begin{equation}\label{eq:ham3bd}
H={p_1^2\over 2m_1}+{p_2^2\over 2m_2} + {(p_1+p_2)^2\over 2m_0} 
- {Gm_0m_1\over r_1} - {G m_0m_2\over r_2} - {Gm_1m_2\over \Delta}
\end{equation}
where $m_0$, $m_1$, $m_2$ are the masses of the star, perturbing primary 
and Trojan planet respectively, $\mathbf{r_1}$, $\mathbf{r_2}$ the 
heliocentric positions of the primary and Trojan planet, $\mathbf{\Delta}
=\mathbf{r_1}-\mathbf{r_2}$ and $\mathbf{p_1}$, $\mathbf{b_2}$ their 
corresponding barycentric momenta. In order to use same units as in the 
ERTBP, one notes that the equations of motion in Cartesian heliocentric 
co-ordinates $\mathbf{r_1}\equiv(r_{1x},r_{1y})$, $\mathbf{r_2}\equiv
(r_{2x},r_{2y})$, and barycentric velocities $\mathbf{p}_1/m_1\equiv
(v_{1x},v_{1y})$, $\mathbf{p}_2/m_2\equiv(v_{2x},v_{2y})$ only depend 
on the variables $r_{ix},r_{iy},v_{ix},v_{iy}$, $i=1,2$ and on the constants 
$Gm_0$, $Gm_1$, $Gm_2$. Then, we solve $Gm_0+Gm_1=1$, $Gm1=\mu$ and assign 
a value to $Gm_2 = Gm_0(m_2/m_0)$ according to the considered mass ratio 
$\mu_2 = m_2/m_0$. 

In order now to obtain comparable FLI maps in the two problems, we
proceed as follows. For every point on the plane of the stability map
of the ERTBP (such as in Fig.~\ref{fig:full3BP-1to6.png}a), we compute
the corresponding heliocentric positions and velocities of both the
primary and the Trojan, i.e. $(r_{ix}(t=0)$, $r_{iy}(t=0)$,
$\dot{r}_{ix}(t=0)$ and $\dot{r}_{iy}(t=0)$, with $i=1,2$. From
Hamilton's equations of~(\ref{eq:ham3bd}) one readily sees that the
barycentric velocities $(v_{ix},v_{iy})$, $i=1,2$ of both bodies are
linear functions of the heliocentric ones. Thus, from every point of
the FLI map in the ERTBP, we compute the complete set of corresponding
initial conditions $r_{ix}(t=0)$, $r_{iy}(t=0)$, $v_{ix}(t=0)$ and
$v_{iy}(t=0)$ needed in order to integrate the full Three Body
problem. Via the same process we assign also corresponding initial
conditions for the variational equations of motion in the two
problems.

Figure \ref{fig:full3BP-1to6.png} shows the comparison of the FLI
stability maps in the case of dominance of the 1:6 resonance, at
$\mu=0.0041$ as $\mu_2$ evolves, i.e., $\mu_2 = 3\times 10^{-6}$ (1
Earth mass, upper row), or $\mu_2 = 3\times 10^{-5}$ (10 Earth masses,
lower row). The left, middle and right panels correspond to initial
eccentricities of the primary equal to $e'=0.02$, $e'=0.06$ and
$e'=0.1$. Thus, these maps are comparable with the ones under the
ERTBP (Fig.~\ref{fig:1to6panels.png}). The main observation is that
switching on the mass $m_2$ results in a considerable reduction of the
area occupied by the stable domains of the secondary resonances. This
is mostly caused by the secular variations induced on the orbit of the
primary planet, which increase the amplitude of modulation of the
separatrices of each secondary resonance. However, the main domain of
stability remains nearly unaffected by these phenomena, and retains a
quite similar width in all simulations with different masses
$\mu_2$. We only see some transverse resonances penetrating the
lowermost (with respect to the eccentricities) part of the stability
map for $\mu_2$ as large as 10 Earth masses. On the other hand, the
analytical determination of the innermost separatrix via the resonant
integral of the `basic model' yields an estimate of the border of the
main stability domain which remains robust against the increase of
$\mu_2$.

\section{Conclusions}\label{sec:concl}

In the present work, we discussed a new application for the `basic Hamiltonian
model'  $H_b$ for Trojan motions presented originally in~\cite{PaezEfthy2015}: 
this is the determination of the border of effective stability, using the 
theoretical separatrices of the most conspicuous secondary resonances of 
$H_b$. In detail: 

1) We compute resonant normal forms for various secondary resonances of 
$H_b$, using an `asymmetric expansion' for the Hamiltonian (see Section 2), 
which allows to speed up the convergence of both the original polynomial 
representation of the Hamiltonian as well as its normal form. The 
improvement obtained by the asymmetric expansion is demonstrated with 
numerical examples. 

2) Using the classical normal form construction with Lie series in 
order to compute a resonant normal form for a specific secondary 
resonance, one ends with an expression for an invariant of the 
normal form called the `resonant integral' $\Psi$ (see Section 3). 
The level curves of $\Psi$ allow, in turn, to obtain a theoretical 
phase portrait on a surface of section, and in particular to compute 
theoretical separatrices as well as the center of the secondary 
resonance.  

3) The method typically yields underestimates of the position of the 
center and outer separatrix of the resonance, but a very efficient 
determination of the inner (closer to the libration center) separatrix 
of the resonance. 

4) We argued that the inner limit $\Delta u_{min}(e_{p,0})$ found in this 
way represents a clear border which exists in numerical stability maps 
between two well distinct domains in the space of the proper elements 
$(\Delta u,e_{p,0})$ (see Section 4 for definitions). The inner domain 
is populated by regular orbits and isolated resonances with regular or 
marginally chaotic orbits, while the outer domain hosts either closely 
packed secondary resonances or a strongly chaotic domain. In fact, with 
increasing value of the primary's eccentricity $e'$, a modulation 
mechanism essentially wipes out all the resonances, creating a large 
outer domain of strong chaos. As a consequence, we argued that the 
inner domain, delimited by the innermost theoretical separatrix of 
the most conspicuous secondary resonance of $H_b$ practically coincides 
with the limit of the effective stability domain for Trojan motions. 

 5) We demonstrated that the role of the secondary resonances of
  the basic model $H_b$, as delimiters of the domain of effective
  stability, covers most of the values of the parameters entering the
  problem (primary's mass and eccentricity, Trojan body's
  eccentricity), while it remains robust even in the full Three Body
  problem, for Trojan bodies of mass $\sim 1$ Earth mass.

\vspace{0.5cm}
\noindent
{\bf Acknowledgements:} Useful discussions with Prof. U. Locatelli
are gratefully acknowledged. R.I.P. was supported by the Research 
Comittee of the Academy of Athens, under the grant 200/854.

\section*{Appendix A}\label{App:variables}
The variables corresponding to the three degrees of freedom appearing in the
expression of the basic Hamiltonian $H_b$ in Eq.\eqref{eq:hbasic}, 
$(u,v)$, $(Y_f,\phi_f)$ and $(Y_p,\phi_p)$ are given in terms of the orbital
elements as follows:

\begin{equation}
u = \lambda - \lambda' - \frac{\pi}{3}~~,
\end{equation}

\begin{equation}
v = \sqrt{a} - 1~~,
\end{equation}

\begin{displaymath}
\beta = \omega - \phi'~~, 
\end{displaymath}

\begin{displaymath}
y = \sqrt{a} \left( \sqrt{1-e^2} -1 \right)~~,
\end{displaymath}

\begin{displaymath}
V = \sqrt{-2y} \sin \beta - \sqrt{-2y_0} \sin \beta_0~~,
\end{displaymath}

\begin{displaymath}
W = \sqrt{-2y} \cos \beta - \sqrt{-2y_0} \cos \beta_0~~,
\end{displaymath}

\begin{displaymath}
Y = - \left( \frac{W^2 + V^2}{2} \right)
\end{displaymath}

\begin{equation}
\phi = \arctan \left( \frac{V}{W} \right) 
\end{equation}

\begin{equation}
\phi_f = \lambda' - \phi~~,
\end{equation}

\begin{equation}
Y_f = \int \frac{\partial E}{\partial \lambda'} \mathrm{d} t + v~~,
\end{equation}

\begin{equation}
Y_p = Y - Y_f~~,
\end{equation}
where $\lambda$, $\omega$, $a$ and $e$
are the mean longitude, the longitude of the perihelion,
the major semiaxis and eccentricity of the Trojan body, 
$\lambda'$ and $\phi' = \omega'$ are the
mean longitude and longitude of the perihelion of the
perturber, $\beta_0 = \pi/3$, $y_0 = \sqrt{1-e'^2} -1$,
and $E$ represents the total energy of the Trojan as 
computed from Eq.~\eqref{eq:h_rmpp} (see \cite{PaezEfthy2015} for
further details in the construction).

\section*{Appendix B}\label{App:assymexp}

The asymmetric expansion in terms of $u=\tau-\pi/3$, up to a generic order $K$
for the functions $\frac{\cos\tau}{(2-2 \cos \tau)^{N/2}}$,
$\frac{\sin\tau}{(2-2 \cos \tau)^{N/2}}$, $\cos^M \tau $ and 
$\sin^M \tau $, with $N,\,M \in \mathbb{N}$ fixed
is given by 
{\small
\begin{displaymath}
\frac{\cos\tau}{(2-2 \cos \tau)^{N/2}} = \frac{1}{2^{N/2}} \, 
\sum_{k=0}^{K} {\cal M}_{1}(k) \, u^k + {\cal O}(u^K)~,\quad \mathrm{where} 
\quad
{\cal M}_{1}(k) =  \sum_{i=k}^{K} \frac{1}{i!} \,F^{(i)}(\pi/2)\,
{i \choose k}
 \left(-\frac{\pi}{6} \right)^{i-k}~~,
\end{displaymath}
\begin{displaymath}
\frac{\sin\tau}{(2-2 \cos \tau)^{N/2}} = \frac{1}{2^{N/2}} \, 
\sum_{k=0}^{K} {\cal M}_{2}(k) \, u^k + {\cal O}(u^K)~,\quad 
\mathrm{where} \quad 
{\cal M}_{2}(k) =  \sum_{i=k}^{K} \frac{1}{i!} \,G^{(i)}(\pi/2)\,
{i \choose k}
 \left( 
-\frac{\pi}{6} \right)^{i-k}~~,
\end{displaymath}
\begin{displaymath}
\cos^M\tau =  \sum_{k=0}^{K} {\cal M}_{3}(k) \, u^k~ + {\cal O}(u^K), 
\quad \mathrm{where} \quad
{\cal M}_{3}(k) = \sum_{i=k}^{K} \frac{1}{i!} \,B_{M,M}^{(i)}\,
{i \choose k}
 \left( -\frac{\pi}{6} \right)^{i-k}~~,
\end{displaymath}
\begin{displaymath}
\sin^M\tau =  \sum_{k=0}^{K} {\cal M}_{4}(k) \, u^k~ + {\cal O}(u^K), 
\quad \mathrm{where} \quad
{\cal M}_{4}(k) = \sum_{i=k}^{K} \frac{1}{i!} \,C_{M,M}^{(i)}\,
{i \choose k} 
 \left( -\frac{\pi}{6} \right)^{i-k}~~,
\end{displaymath}
}
and
{\small
\begin{align*}
F^{(n)}(\pi/2) &= \sum_{i=1}^{[\frac{n-1}{2}]} (n,2i-1)\, (-1)^{i} 
\,f^{(n-(2i-1))}(\pi/2)~~, \\
G^{(n)}(\pi/2) &= \sum_{i=0}^{[\frac{n}{2}]} (n,2i)\, (-1)^{i} 
\,f^{(n-2i)}(\pi/2)~~,
\end{align*}
}
with $[\frac{n-1}{2}]$ the integer part of $\frac{n-1}{2}$, and
$[\frac{n}{2}]$ the integer part of $\frac{n}{2}$; the derivatives
$f^{(n)}$ are given by
\begin{displaymath}
f^{(n)}\left( \pi/2 \right) = \sum_{m=1}^{n} A_{m,m}^{(n)} ~~;
\end{displaymath}
the coefficients $A_{m,m}^{(n)}$, $B_{M,M}^{(n)}$ and $C_{M,M}^{(n)}$ 
are given by
\begin{align*}
A_{m,m}^{(n)} &= -A_{m,m-1}^{(n-1)} - \left( \frac{2(m-1)+N}{2} \right) 
A_{m-1,m-1}^{(n-1)}~, 
 &A_{1,1}^{(1)} &= - \frac{N}{2}~~, \\
B_{M,M}^{(n)} &= - B_{M,M-1}^{(n-1)} + \left(M+1\right) B_{M,M+1}^{(n-1)},
 &B_{1,1}^{(1)} &= -M~~, \phantom{\Big( A \Big)} \\
C_{M,M}^{(n)} &= C_{M,M-1}^{(n-1)} - \left(M+1\right) C_{M,M+1}^{(n-1)},
 &C_{1,1}^{(1)} &= M~~. \phantom{\Big( A \Big)}
\end{align*}
For a proof of these formul\ae, we refer the reader to~\cite{thesis}.



\end{document}